\documentclass[aps,prb,showpacs,superscriptaddress,groupedaddress]{revtex4-1}
\usepackage{graphicx}
\usepackage{amsmath}
\usepackage{amssymb,color}
\usepackage[utf8]{inputenc}
\usepackage{bbm}
\usepackage[markup=nocolor]{changes}
\usepackage{braket}
\input{epsf}
\newcommand{\nn}{\nonumber}
\newcommand{\kk}{\textbf{k}}
\newcommand{\qv}{\textbf{q}}
\newcommand{\rr}{\textbf{r}}

\newcommand{\ua}{\uparrow}
\newcommand{\da}{\downarrow}

\begin{document}

	\title{Quasiparticle Interference and Symmetry of Superconducting Order Parameter in Strongly Electron-Doped Iron-based Superconductors}
	
	\author{Jakob~B\"oker$^1$}
	\author{Pavel~A. Volkov$^{2,1}$}
	\author{P. J. Hirschfeld$^3$}
	\author{Ilya~Eremin$^1$}
	\affiliation{1-Institut f\"ur Theoretische Physik III, Ruhr-Universit\"at Bochum, D-44801 Bochum, Germany}
	\affiliation{2-Department of Physics and Astronomy, Center for Materials Theory, Rutgers University, Piscataway, New Jersey 08854, USA}
	\affiliation{3-Department of Physics, University of Florida, Gainesville, Florida 32611, USA}
	
	\begin{abstract}
		Motivated by recent experimental reports of significant spin-orbit coupling (SOC) and a sign-changing order-parameter
		in the Li$_{1-x}$Fe$_x$(OHFe)$_{1-y}$Zn$_y$Se superconductor with only electron pockets present, we study the possible Cooper-pairing symmetries and their quasiparticle interference (QPI) signatures. We find that each of the resulting states - $s$-wave, $d$-wave and helical $p$-wave - can have a fully gapped density of states (DOS) consistent with angle-resolved photoemission  spectroscopy (ARPES) experiments and, due to spin-orbit coupling, are a mixture of spin singlet and triplet components leading to intra- and inter-band features in the QPI signal. Analyzing predicted QPI patterns we find that only the spin-triplet dominated even parity $A_{1g}$ (s-wave) and $B_{2g}$ (d-wave) pairing states  are consistent with the experimental data. Additionally, we show that these states can indeed be realized in a microscopic model with atomic-like interactions and study their possible signatures in spin-resolved STM experiments. 
		
	\end{abstract}
	\date{\today}
	
	\maketitle
	
	\section{Introduction}
	
	In iron-based superconductors, it has been widely believed that superconductivity is driven by repulsive interactions, enhanced by the presence of the spin 
	fluctuations associated with the parent antiferromagnetic state.  In this scenario, these fluctuations drive a sign
	reversal ($s_\pm$ state) \cite{Kuroki2008,Mazin2008,Chubukov2008,Chubukov_review2012,Hirschfeld_review2016} between order parameters (OP) on the electron and hole
	Fermi surface pockets at the $M$- and $\Gamma$-point, respectively. The discovery of superconductivity in intercalated or
	monolayer FeSe at a critical temperature of the order above 40K revived  
	interest in Fe-based superconductivity, but raised further questions on the origin of superconductivity in these compounds\cite{Guo10,Lu15,Wang12,Miyata15,Zhao16},  because, unlike bulk FeSe, ARPES experiments show that many of these  FeSe-derived systems appear to be missing the 
	hole pockets at the $\Gamma$-point required in the the conventional scenario. 	

	Initial model calculations  based  on the multiorbital spin-fluctuation framework for systems manifesting only electron pockets at the $M$-point predicted  $d$-wave  symmetry state in this case\cite{Meier11,Wang11},  driven by the  spin fluctuations  connecting the electron pockets that remain when the  hole pockets are removed. In the proper 2-Fe unit cell, such a state must have  gap nodes on the Fermi surface\cite{Mazin_dwavenode_2011}. This is because the electron pockets located near $(\pi,0)$ or $(0,\pi)$ points of the Brillouin Zone (BZ) in the 1-Fe unit cell fold onto $(\pi,\pi)$ point of the folded BZ as the crystallographic symmetry lowers due to
	the Se positions. This may lead to hybridization between the electron pockets\cite{Khodas12}, which then forces  the $d_{x^2-y^2}$-state to acquire gap nodes, although in principle the nodal area may
	be very small,  proportional to the hybridization (``quasinodes"). On the other hand,  ARPES experiments in most of the electron-intercalated materials indicated a nodeless superconducting (sc) state\cite{Yan16,Zhang16}. Several proposals  for the gap structure have been put forward, including a conventional $s^{++}$-wave scenario based on the electron-phonon interaction and orbital fluctuations\cite{Onari12}, as well as the ``bonding-antibonding" scenario\cite{Mazin_dwavenode_2011,HKM_ROPP} in which  the order parameter on the inner electron pocket (mostly $d_{xz}/d_{yz}$ character)  has one sign, and on the outer electron pocket (mostly $d_{xy}$ character) the other\cite{Khodas12}.
	Furthermore, it has been argued that the hybridization of the electron pockets is mainly due to SOC\cite{Kreisel2013,Cvetkovic13,Myles18}, which within a 3D spin fluctuation framework may stabilize the bonding-antibonding $s_\pm$ state against d-wave\cite{Kreisel2013} and mixes  a spin-triplet component into the even parity $s^{+-}$-wave state\cite{Cvetkovic13,Myles18}. Overall, one can see that the sign structure of the superconducting order parameter is intimately related to the pairing mechanism. Therefore, experiments allowing to determine it could be of great potential importance.
	
	One rapidly developing technique to determine the phase structure of the order parameter makes use of QPI as measured by Fourier transform scanning tunneling microscopy (FT-STM). This probe measures the wavelengths
	of Friedel oscillations caused by impurities present in a metallic or superconducting system, which in turn contains information on the electronic structure of the pure system. A
	subset of  scattering wave vectors {\bf q} corresponding to peaks in the FT-STM can be enhanced or
	not according to the type of disorder and the phase structure
	of the superconducting gap\cite{Franz03,Nunner06}. Recently it was proposed  by Hirschfeld, Altenfeld, Eremin and Mazin (HAEM)\cite{Hirschfeld(2015)} that the sign structure of the order parameter in a multiband system can
	be extracted from the Fourier transform of the local density of states QPI pattern near an impurity in the
	superconducting state. The antisymmetrized QPI intensity integrated over the wavevectors corresponding to scattering between
	two bands was shown to have a dependence on frequency very different for sign-changing and sign preserving scenarios leading
	to a strong, single-sign enhancement of the integrated response in the former case. This  qualitative result was also confirmed by extensive numerical simulations with finite disorder\cite{Martiny17}.   Recently, a complementary phase sensitive technique to detect sign-changing gaps in the presence of strong impurity bound states was proposed\cite{Chi2017}.

	Using QPI analysis, the authors of Ref. \onlinecite{Sprau2017} were able to identify a sign changing order parameter
	in FeSe. Most importantly  for our purposes, similar conclusions were recently drawn for the strongly electron doped iron-based superconductor lithium
	hydroxide intercalated FeSe\cite{Du2017}. In other words, the order parameter in Li$_{1-x}$Fe$_x$(OHFe)$_{1-y}$Zn$_y$Se, alternates sign, either between the Fermi
	surface sheets, or within individual sheets.  However, distinguishing between these alternatives was beyond the resolution of the experiment. In any case,  the situation is somewhat more  complicated than anticipated in Ref. \onlinecite{Du2017}, since  the effect of spin-orbit interaction on pairing needs to be taken into account as well. Moreover,  recent observation of  Majorana zero modes in (Li$_{0.84}$Fe$_{0.16}$)OHFeSe\cite{Liu2018} suggests possible  broader implications of the spin-orbit coupling for the Cooper-pairing in electron doped intercalated iron-based superconductors. Note that in contrast to  Ref. \onlinecite{Du2017}, no Zn substitution was used in Ref. \onlinecite{Liu2018}. The amount of Zn, however, is relatively small (less than 2 percent). This amount does not affect the superconducting transition temperature or the electronic structure in a significant way and is done only for the purpose of enhancing the QPI signal in the scanning tunneling microscopy.

	In this manuscript we study the possible Cooper-pairing symmetries  and their QPI signatures for  strongly electron-doped Fe-based superconductors using the effective three-orbital model of Refs. \onlinecite{Cvetkovic13,Myles18} with spin-orbit coupling and proper consideration of all lattice symmetries of the FeSe space group. We find that each of the resulting states - $A_{1g}$-wave, $B_{2g}$-wave and helical $E_{u}$-wave - can have a fully gapped DOS consistent with  ARPES experiments and, due to spin-orbit coupling, are a mixture of spin singlet and triplet components leading to intra- and inter-band features in the QPI signal. Analyzing predicted QPI patterns we find that $A_{1g}$-wave pairing state, with the two dominant peaks in the  DOS  roughly  corresponding  to  the  gap  energies  on each pocket, and $B_{2g}$-wave pairing state both with a significant even parity spin triplet component are consistent with  the  experimental  data. Moreover, we show that pairing states with dominant spin triplet component can be identified  using spin-resolved STM.

	\section{Model}
	\begin{figure}[h]
		\centering
		\includegraphics[width=8.2cm,keepaspectratio]{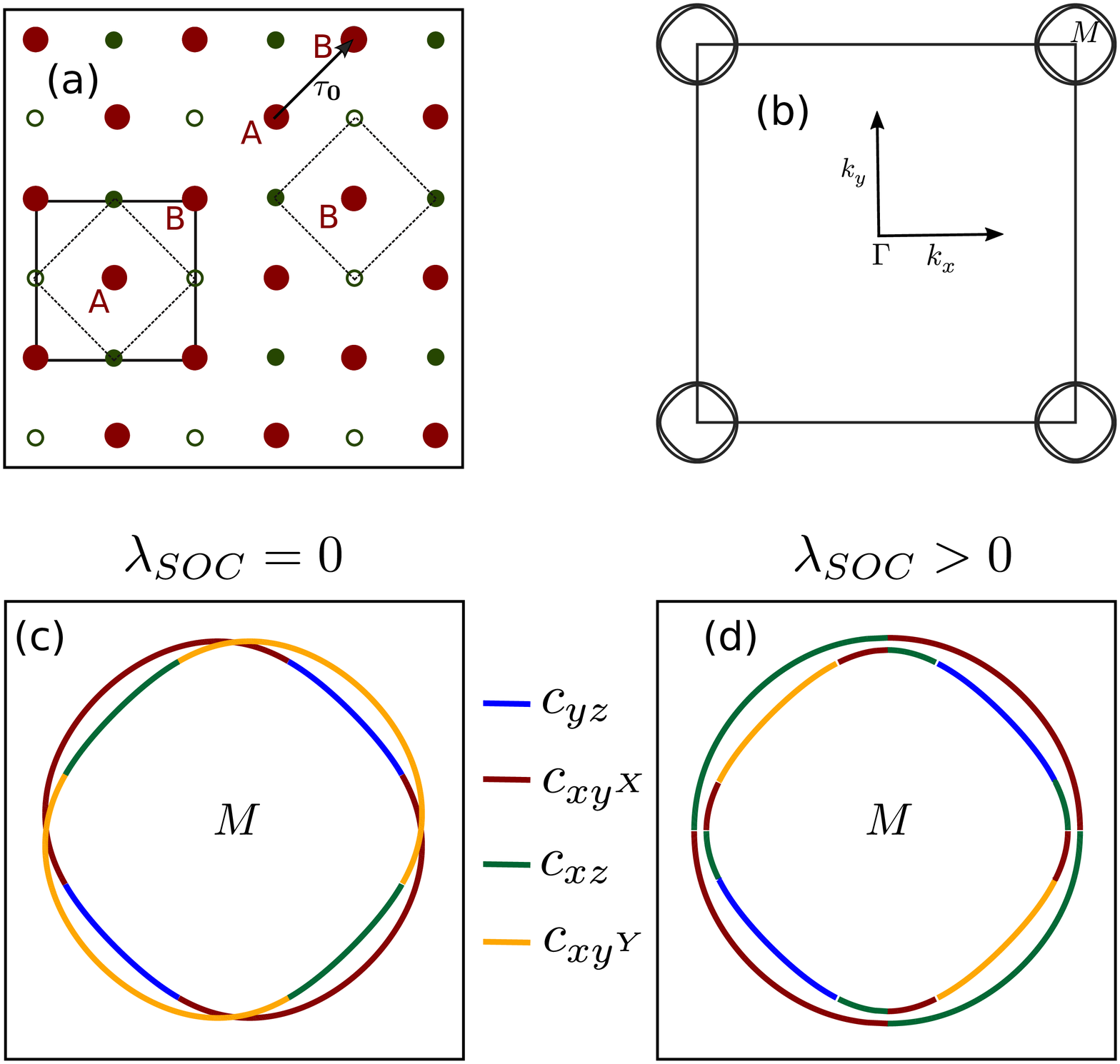}	
		\caption{(a) Single layer of the iron based superconductors lattice structure. Red and green dots are iron and pnictogen or Se atoms, respectively. One pnictogen sublattice is puckered above the iron layer (filled green dots) one is puckered bellow (empty green dots) which divides the iron atoms into sublattices A and B. One-iron unit cells for sublattices A and B are denoted by dashed squares. The two-iron unit cell, taking the puckering into account, is shown by the solid square. The vector $\mathbf{\tau_0}$ connects sublattices A and B. (b) Fermi surface of low-energy model consistent with Ref.\onlinecite{Zhao16}. (c) Pockets at M point without SOC corresponding to is X- and Y-pocket folded upon each other. (d) Pockets at M point with SOC $\lambda_{\text{SOC}}=5$ meV. Inner and outer pocket in the presence of SOC, leading to lifted degeneracy at zone diagonals.  The color scheme in (c),(d) follows majority orbital content.}	
		\label{Fig:Lattice}
	\end{figure}
	We wish to describe the low energy states near the M,-points of the Brillouin zone using the orbitally projected band model of Ref. \onlinecite{Cvetkovic13,Myles18} for the two-iron unit cell. Near the Fermi level, only the $xz$, $yz$ and $xy$ orbitals  contribute significantly; hence, the full 10-orbital tight binding model  is projected onto the subspace of these three orbitals. 
	The effective low energy Hamiltonian near the M-point that takes into account all the lattice symmetries of the FeSe space group as well as time reversal symmetry is defined as
	\begin{eqnarray}
	H=\sum_{\sigma,\sigma^\prime=\ua,\da}\sum_\kk &\Psi^\dagger_{M,\sigma}(\kk)
	\left(\begin{array}{cc}
	h^{\prime\sigma,\sigma^\prime}_{X}(\kk) & \Lambda^{\sigma,\sigma^\prime}_{\text{SOC}} \\ 
	\Lambda^{\dagger\sigma,\sigma^\prime}_{\text{SOC}} & h^{\prime\sigma,\sigma^\prime}_{Y}(\kk)
	\end{array} \right)
	\Psi_{M,\sigma^\prime}(\kk),\label{Eq:Model}\nn\\
	\end{eqnarray}
	where the four component spinor $\Psi^\dagger_{M,\sigma}(\kk)=\left(\Psi^\dagger_{X,\sigma}(\kk),\Psi^\dagger_{Y,\sigma}(\kk)\right)$ describes the states at the M-point for each spin projection $\sigma$. The doublets $\Psi_{X,\sigma}(\kk)$ and $\Psi_{Y,\sigma}(\kk)$ are defined as
	\begin{align}
	\Psi_{X,\sigma}(\kk)=
	\left(\begin{array}{c}
	c_{yz,\sigma}(\kk) \\ 
	c_{xy^X,\sigma}(\kk)
	\end{array} \right)\quad,\quad \Psi_{Y,\sigma}(\kk)=\left(\begin{array}{c}
	c_{xz,\sigma}(\kk) \\ 
	c_{xy^Y,\sigma}(\kk)
	\end{array} \right)
	\label{XadYSpinor}
	\end{align}
	Moreover, we have
	\begin{align}
	h^{\prime\sigma\sigma^\prime}_X(\mathbf{k})&=h_X(\mathbf{k})\delta_{\sigma,\sigma^\prime}+\Big(\lambda_z(k_x-k_y)\nn\\
	&+p_{z_1}(k_x^3-k_y^3)+p_{z_2}k_xk_y(-k_x+k_y)\Big)\sigma^z_{\sigma,\sigma^\prime}\tau_1\nn\\
	h^{\prime\sigma\sigma^\prime}_Y(\mathbf{k})&=h_Y(\mathbf{k})\delta_{\sigma,\sigma^\prime}+\Big(\lambda_z(k_x+k_y)\nn\\
	&+p_{z_1}(k_x^3+k_y^3)+p_{z_2}k_xk_y(k_x+k_y)\Big)\sigma^z_{\sigma,\sigma^\prime}\tau_1\nn\\\label{InterBandSoC}
	\end{align}
	and
	\begin{align}
	h_X(\mathbf{k})
	&=\left(\begin{array}{cc}
	\epsilon_1+\frac{\mathbf{k}^2}{2m_1}+\alpha_1k_xk_y & -iv(k_x+k_y) \\ 
	iv(k_x+k_y) & \epsilon_3+\frac{\mathbf{k}^2}{2m_3}+\alpha_3k_xk_y
	\end{array} \right)\label{h_X},\\
	h_Y(\mathbf{k})
	&=\left(\begin{array}{cc}
	\epsilon_1+\frac{\mathbf{k}^2}{2m_1}-\alpha_1k_xk_y & -iv(-k_x+k_y) \\ 
	iv(-k_x+k_y) & \epsilon_3+\frac{\mathbf{k}^2}{2m_3}-\alpha_3k_xk_y\label{h_Y}
	\end{array} \right)	
	\end{align}
	where the Pauli matrices $\{\sigma^x,\sigma^y,\sigma^z\}$ and  $\{\tau_1,\tau_2,\tau_3\}$ act on spin and orbital space, respectively. The $\lambda_z$, $p_{z_1}$ and $p_{z_2}$ terms in eq.(\ref{InterBandSoC}) describe the k-dependent intra-band SOC which does not couple the two Fermi pockets but lifts the out-of plane spin degeneracy. The inter-band SOC term which hybridizes X- and Y-pocket  is given by
	\begin{align}
	\Lambda^{\sigma,\sigma^\prime}_{\text{SOC}}=i\lambda_{\text{SOC}}\left(\frac{\tau_1+i\tau_2}{2}\otimes\sigma^x_{\sigma,\sigma^\prime}+\frac{\tau_1-i\tau_2}{2}\otimes\sigma^y_{\sigma,\sigma^\prime}\right)\label{Eq:SOC}.
	\end{align}
	In order to describe intercalated FeSe we use the Luttinger invariants, Tab.(\ref{Tab:Fitting parameterIntercalated}), which were evaluated in Ref.{\onlinecite{Myles18}} based on the available ARPES data. 
	\begin{table}[h]
		\begin{tabular}{c c c c}
			\hline 
			$\alpha_1$ & & & $782.512 \enspace\text{meV}\mathring{A}^2$ \\ 
			
			$\alpha_3$ & & & $-1400\enspace\text{meV}\enspace\mathring{A}^2$ \\ 
			
			$\frac{1}{2m_1}$ & & & $-492.01\enspace\text{meV}\enspace\mathring{A}^2$ \\ 
			
			$\frac{1}{2m_3}$ & & & $1494.14\enspace\text{meV}\enspace\mathring{A}^2$ \\ 
			
			$v$ & & & $224.406\enspace\text{meV}\enspace\mathring{A}$ \\ 
			$\lambda_z$ & & & $26\enspace\text{meV}$ \\
			$p_{z_1}=p_{z_3}$ & & & $0$ \\
			\hline 
		\end{tabular} 
		\caption{Fitting parameters relevant for intercalated FeSe taken from Ref.{\onlinecite{Myles18}}. }
		\label{Tab:Fitting parameterIntercalated}
	\end{table}
	\vskip .2cm
	The value of the intraband spin-orbit coupling, $\lambda_z$ is in agreement with those found in ab-initio calculations and ARPES experiments\cite{Borisenko2016,Day2018}. Furthermore, the value of the interband spin-orbit coupling, $\lambda_{\text{SOC}}$ between the electron pockets separated by the large momentum yielding their hybridization and splitting on the Fermi surface is found to be smaller and is taken to be $\sim 5$ meV\cite{Day2018}.   
	\vskip .2cm
	Diagonalizing Eq.(\ref{Eq:Model}) yields four bands: two regular ones that form the inner and outer electron pocket, see Fig.\ref{Fig:Lattice}(b), and two incipient bands that do not cross the Fermi level. The effect of inter-band SOC on the band structure is visualized in Fig.\ref{Fig:Lattice}(c) and Fig.\ref{Fig:Lattice}(d).

	\section{Mean field phase diagram}
	
	Although phenomenologically the classification of superconducting orders for two electron pockets was considered previously\cite{Cvetkovic13,Myles18} we analyze here its microscopic formulation via mean-field treatment of the  atomic on-site interactions given by the Hubbard and Hund's couplings $U$, $U^\prime$, $J$ and $J^\prime$ which enter the Hubbard-Kanamori Hamiltonian as
	\begin{align}
	H_{\text{int}}(\rr)=\sum_{\alpha}&\sum_{\mu} U_{\mu\mu}   d^{\alpha\dagger}_{\mu,\ua}(\textbf{r})d^{\alpha\dagger}_{\mu,\da}(\textbf{r})d^{\alpha}_{\mu,\da}(\textbf{r})d^{\alpha}_{\mu,\ua}(\textbf{r})\nn\\
	+ &\sum_{\mu\neq \nu} J_{\mu\nu}^\prime  d^{\alpha\dagger}_{\mu,\ua}(\textbf{r})d^{\alpha\dagger}_{\mu,\da}(\textbf{r})d^{\alpha}_{\nu,\da}(\textbf{r})d^{\alpha}_{\nu,\ua}(\textbf{r})\nn\\
	+ &\sum_{\mu<\nu} \sum_{\sigma,\sigma^\prime} \frac{J_{\mu\nu}}{2}  d^{\alpha\dagger}_{\mu,\sigma}(\textbf{r})d^{\alpha\dagger}_{\nu,\sigma^\prime}(\textbf{r})d^{\alpha}_{\mu,\sigma^\prime}(\textbf{r})d^{\alpha}_{\nu,\sigma}(\textbf{r})\nn\\
	+ &\sum_{\mu<\nu} \sum_{\sigma,\sigma^\prime} \frac{U_{\mu\nu}^\prime}{2} d^{\alpha\dagger}_{\mu,\sigma}(\textbf{r})d^{\alpha\dagger}_{\nu,\sigma^\prime}(\textbf{r})d^{\alpha}_{\nu,\sigma^\prime}(\textbf{r})d^{\alpha}_{\mu,\sigma}(\textbf{r}).\label{InteactionHamiltonian}
	\end{align}
	Here $\alpha\in\{A,B\}$, $\{\sigma,\sigma^\prime\}\in\{\ua,\da\}$ and $\{\mu,\nu\}\in\{yz,xz,xy\}$ label lattice sites, spins and orbitals, respectively. $d^{\alpha\dagger}_{\mu,\sigma}(\rr)$ and $d^{\alpha}_{\mu,\sigma}(\rr)$ are the second quantized operators creating and annihilating particles on sub-lattice A and B, see Fig.\ref{Fig:Lattice}(a). Using the results presented in Ref. \onlinecite{Cvetkovic13} and assuming sharply localized Wannier functions of the $xz$, $yz$ and $xy$ orbitals, we can relate, up to a constant, the $d^{\alpha}_{\mu,\sigma}(\kk)$ operators acting in the one iron unit cell to the components of the doublets $\Psi_{X,\sigma}(\kk)$ and  $\Psi_{Y,\sigma}(\kk)$ in the two iron unit cell via
	
	\begin{align}
	d^{A(B)}_{xz,\sigma}(\kk)&\propto \pm\frac{1}{\sqrt{2}}c_{xz,\sigma}(\kk),\label{Eq:Projection1}\\
	d^{A(B)}_{yz,\sigma}(\kk)&\propto \frac{1}{\sqrt{2}} c_{yz,\sigma}(\kk),\label{Eq:Projection2}\\
	d^{A(B)}_{xy,\sigma}(\kk)&\propto\frac{1}{\sqrt{2}}\left(c^{X}_{xy,\sigma}(\kk)\pm c^{Y}_{xy,\sigma}(\kk)\right),\label{Eq:Projection3}
	\end{align}
	where  we absorb the constant prefactors into the Hubbard and Hund terms. As the $xy$ orbital contributes to both X- and Y-point eq.(\ref{Eq:Projection3}) leads to "Umklapp" terms at the M-point. 
	
	We assume that even parity solutions are still the leading pairing instabilities (for odd parity solutions see Appendix \ref{sec_A2}) and use Eqs.(\ref{Eq:Projection1}-\ref{Eq:Projection3}) to project eq.(\ref{InteactionHamiltonian}) onto the low energy model decoupled into the spin singlet $A_{1g}$ $s$-wave and $B_{2g}$ $d$-wave symmetry states which at the M-point in presence of inter-band SOC couple to the $E_g$ even parity spin triplet state. When defining the two doublets $\Psi^T_{1\sigma}(\kk)=(c_{yz\sigma}(\kk),c_{xz\sigma}(\kk))$ and $\Psi^T_{3\sigma}(\kk)=(c_{xy^X\sigma}(\kk),c_{xy^Y\sigma}(\kk))$  the pairing terms read

	\begin{align}
	&H^{A_{1g}}_{\text{int}}+H^{B_{2g}}_{\text{int}}+H^{E_{g}}_{\text{int}}\nn\\
	&=\sum_{\kk,\qv}
	\Big(\Psi^\dagger_{1\ua}(\kk)\tau_0\Psi^*_{1\da}(-\kk),\Psi^\dagger_{3\ua}(\kk)\tau_0\Psi^*_{3\da}(-\kk)\Big)\times\nn\\
	&\times\left(\begin{array}{cc}
	\frac{1}{2}(U+J_{11}^\prime) & J_{13}^\prime\\ 
	J_{13}^\prime & U \\ 
	\end{array} \right)
	\left(\begin{array}{c}
	\Psi^T_{1\da}(-\qv)\tau_0\Psi_{1\ua}(\qv)\\ 
	\Psi^T_{3\da}(-\qv)\tau_0\Psi_{3\ua}(\qv)
	\end{array} \right)\nn\\
	+&\frac{1}{2}\Big[\Psi^\dagger_{1\ua}(\kk)\tau_3\Psi^*_{1\da}(-\kk)\Big](U-J_{11}^\prime)\Big[\Psi^T_{1\da}(-\qv)\tau_3\Psi_{1\ua}(\qv)\Big]\nn\\
	+&\frac{1}{2}\Big[\Psi^\dagger_{1\sigma}(\kk)\tau_1\Psi_{3\sigma}(-\kk)\Big](U_{13}^\prime-J_{13})\Big[\Psi^T_{3\sigma}(-\qv)\tau_1\Psi_{1\sigma}(\qv)\Big].\label{Interaction}
	\end{align}

	The 2x2 block in eq.(\ref{Interaction}) corresponds to $A_{1g}$ $s$-wave  pairing.  We write $J^\prime_{13}=\alpha J^\prime_{11}$ and find two eigenvalues $E_{A_{1g}}=\frac{1}{4}\left(J_{11}^\prime2+3U\pm\sqrt{(J^\prime_{11})^2+\alpha 16J_{11}^\prime-2J^\prime_{11}U+U^2}\right)$ which correspond to ordinary "plus-plus" ($s^{++}$) $s$-wave and sign-changing "plus-minus" ($s^{\pm}$) $s$-wave pairing, respectively. While the former channel is purely repulsive without spin-orbit coupling the latter becomes attractive once $J^\prime_{11}>(U+U\sqrt{1+8\alpha})/4\alpha$.\newline 
	The paring term that leads to $B_{2g}$ $d$-wave is $E_{B_{2g}}=\frac{1}{2}(U-J_{11}^\prime)$ and can be directly read off. It is attractive once $J_{11}^\prime>U$ and competes with sign-changing s-wave. Since we assume sharply located Wannier functions which yields eqs.(\ref{Eq:Projection1}-\ref{Eq:Projection3}) and on-site interactions only we find that within our simple mean field approximation the $xy$ orbitals do not contribute to $d$-wave  pairing as no "pair-hopping" term $J^\prime_{33}$ mediates between $xy^X$ and $xy^Y$. This changes once the higher-order diagrams (spin fluctuations) are taken into account.

	The $E_g$ even parity spin triplet corresponds to pairing between the first (second) component of $\Psi_{X\sigma}(\kk)$ ($\Psi_{Y\sigma}(\kk)$) and the second (first) component of $\Psi_{Y\sigma}(\kk)$ ($\Psi_{X\sigma}(\kk)$) and is thus inter-band and attractive once $E_{E_g}=U_{13}^\prime-J_{13}<0$. 
	We perform a mean-field decoupling of Eq.(\ref{Interaction}) into $A_{1g}$  and $B_{2g}$ spin singlet channels with the pairing terms given by 
	
	\begin{align}
	A^s_{1g}:\quad&\Delta^A_1\Psi^T_{1,\sigma}(-\kk)\tau_0 i\sigma^y\Psi_{1,\sigma^\prime}(\kk)\\
	&\Delta^A_3\Psi^T_{3,\sigma}(-\kk)\tau_0 i\sigma^y\Psi_{3,\sigma^\prime}(\kk)\\
	B^s_{2g}:\quad &\Delta^B_1\Psi^T_{1,\sigma}(-\kk)\tau_3 i\sigma^y\Psi_{1,\sigma^\prime}(\kk)\\
	&\Delta^B_3\Psi^T_{3,\sigma}(-\kk)\tau_3 i\sigma^y\Psi_{3,\sigma^\prime}(\kk).
	\end{align}
	
	In terms of $\Psi_{M,\sigma}(\kk)$ a triplet term can be written as $\Psi_{M,\sigma}^T(-\kk)\hat{\mathcal{M}}i\sigma^y\boldsymbol{\sigma}\Psi_{M,\sigma}(\kk)$, where $\hat{\mathcal{M}}$ and $i\sigma^y\boldsymbol{\sigma}$ represent orbital and spin part, respectively. Since $i\sigma^y\boldsymbol{\sigma}$ is symmetric an even (odd) parity triplet requires $\hat{\mathcal{M}}$ to be anti-symmetric (symmetric). $i\sigma^y\boldsymbol{\sigma}$ can be divided into an in plane $i\sigma^y(\sigma^x, \sigma^y)$ and out of plane $i\sigma^y\sigma^z$ component which transform as  as the two dimensional $E_g$ and one dimensional $A_{2g}$ irreducible representation, respectively. 
	In presence of SOC, orbital and spin degrees of freedom transform together under operations of the space group.
	We focus on the $E_g$ even parity spin triplet that together with the $E_g$ in-plane spin component decomposes into a direct sum of one dimensional representations as $E_g\otimes E_g=A_{1g}\oplus B_{1g}\oplus B_{2g}\oplus A_{2g}$. Using the two anti symmetric components of $E_g$ and $\tau_{\pm}=(\tau_1\pm i\tau_2)/2$
	\begin{align}
	E_{g1}^-=i\left(\begin{array}{cc}
	& \tau_- \\ 
	-\tau_+ & 
	\end{array} \right) \quad  E_{g2}^-=i\left(\begin{array}{cc}
	& \tau_+ \\ 
	-\tau_- & 
	\end{array} \right)
	\end{align}
	one finds two even parity spin triplets that transform according to $A_{1g}$ and $B_{2g}$ and hence, couple to the singlet channel.

	\begin{align}
	&A^t_{1g}:\Delta^A_t\Psi_{M,\sigma}^T(-\kk)(-E^-_{g1},iE^-_{g2})(\sigma^z,\sigma_o)\Psi_{M,\sigma}(\kk)\\
	&B^t_{2g}:\Delta^B_t\Psi_{M,\sigma}^T(-\kk)(-E^-_{g1},-iE^-_{g2})(\sigma^z,\sigma_o)\Psi_{M,\sigma}(\kk)
	\end{align}
	We refer the reader to Refs.\cite{Kang2016,Myles18} for further details.
	
\begin{figure}[h]
		\centering
		\includegraphics[width=8.2cm,keepaspectratio]{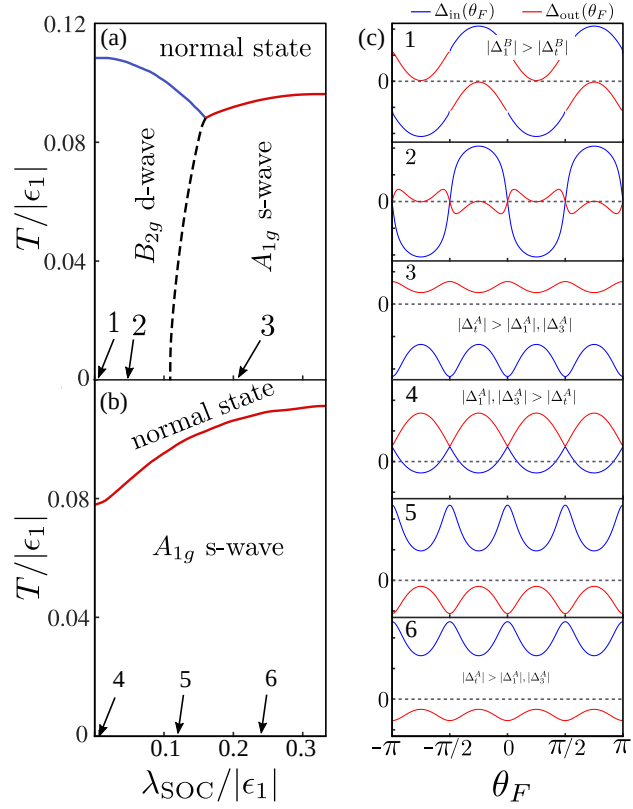}	
		\caption{Phase diagram as function of temperature and inter-band SOC in units of $|\epsilon_1|$. $\epsilon_1=-45$ meV and $\epsilon_2=-95$ meV are energies at the M-point of the two regular and the two incipient bands, respectively. The dimensionless coupling constants ($\tilde{U}=N_0U$ etc. with $N_0=\frac{2|m1|}{2\pi}$) are for (a) $\tilde{U}=0.5$, $\tilde{J}_{11}^\prime=2.58$, $\tilde{J}_{13}^\prime=0.83\tilde{J}_{11}^\prime$, $\tilde{U}^\prime=0.1$ and $\tilde{J}=4.6$ and for (b) $\tilde{U}=0.5$, $\tilde{J}_{11}^\prime=2.58$, $\tilde{J}_{13}^\prime=\tilde{J}_{11}^\prime$, $\tilde{U}^\prime=0.1$ and $\tilde{J}=4.6$. c) Superconducting gap projected on the inner (blue) and outer (red) Fermi-surface as function of the Fermi angle, zero temperature and at three different values of $\lambda_{\text{SOC}}$. }	
		\label{Fig:Phasediagram}
	\end{figure}

		In terms of the spinor $\Psi^T_g(\kk)=(\Psi_{X\ua}(\kk), \Psi_{Y\da}(\kk), \Psi^\dagger_{X\da}(\kk), -\Psi^\dagger_{Y\ua}(\kk))$  the BdG-Hamiltonian reads
	
	\begin{align}
	H_{BdG}^{A_{1g}(B_{2g})}=\sum_{\kk}\Psi^\dagger_g(\kk)\left(\begin{array}{cc}
	\mathcal{H}_0(\kk) & \hat{\Delta}_{A_{1g}(B_{2g})} \\ 
	\hat{\Delta}^\dagger_{A_{1g}(B_{2g})} & -\mathcal{H}_0(\kk)
	\end{array} \right)\Psi_g(\kk)
	\end{align}
	
	with
	
	\begin{align}
	\mathcal{H}_0(\kk)=\left(\begin{array}{cc}
	h^{\prime\ua\ua}_{X}(\kk) & \lambda_{\text{SOC}}\Lambda \\ 
	\lambda_{\text{SOC}}\Lambda^\dagger & h^{\prime\da\da}_{Y}(\kk)
	\end{array} \right)\quad,\quad\Lambda=\left(\begin{array}{cc}
	0 & i \\ 
	1 & 0
	\end{array} \right)
	\end{align}
	and the pairing terms 
	
	\begin{align}
	\hat{\Delta}_{A_{1g}}&=\left(\begin{array}{cccc}
	\Delta^A_1 &  &  & - i\Delta^A_t \\ 
	& \Delta^A_3 & -\Delta^A_t &  \\ 
	& -\Delta^A_t & \Delta^A_1 &  \\ 
	i\Delta^A_t &  &  & \Delta^A_3
	\end{array} \right)\label{s-wave},\\
	\hat{\Delta}_{B_{2g}}&=\left(\begin{array}{cccc}
	\Delta^B_1 &  &  &  i\Delta^B_t \\ 
	& \Delta^B_3 & -\Delta^B_t &  \\ 
	& -\Delta^B_t & -\Delta^B_1 &  \\ 
	- i\Delta^B_t &  &  & -\Delta^B_3
	\end{array} \right).\label{d-wave}
	\end{align}
	where the gaps in orbital space are given by the equations (\ref{Eq:s-wave}) and (\ref{Eq:d-wave}). We self-consistently compute $\hat{\Delta}_{A_{1g}}$ and $\hat{\Delta}_{B_{2g}}$ as a function of temperature and inter-band SOC for the two cases $E_{A1g}<E_{B2g}$ and $E_{A1g}>E_{B2g}$ and present the phase diagrams in Fig.\ref{Fig:Phasediagram}(a) and Fig.\ref{Fig:Phasediagram}(b), respectively. Figure\ref{Fig:Phasediagram}(c) shows the angular dependence of the superconducting gap projected onto inner (blue) and outer (red) Fermi surface for three values of $\lambda_{\text{SOC}}$ marked by black arrows. The gap projected on the l'th band is given by $\Delta_{l}(\kk)=\sum_{\alpha,\beta}\mu^\dagger_{l\alpha}(\kk)\hat{\Delta}_{\alpha\beta}\mu_{\beta l}(\kk)$ where $\hat{\mu}(\kk)$ diagonalizes $\mathcal{H}_0(\kk)$.\newline 
	
	In the case when solely $E_{A_{1g}}$ ($E_{B_{2g}}$) is attractive, $\Delta^{A}_t$ ($\Delta^{B}_t$) is induced by SOC and scales with $\lambda_{\text{SOC}}$. In this case $|\Delta^{A(B)}_t|<|\Delta^{A(B)}_1|, |\Delta^{A(B)}_3|$ and we call the state a \emph{spin singlet dominated} $A_{1g}$ ($B_{2g}$). If $E_{E_g}$ is attractive, however,  $\Delta_t$  can develop independently of $\Delta_{1,3}$ and allows for states where $|\Delta^{A(B)}_t|>|\Delta^{A(B)}_1|, |\Delta^{A(B)}_3|$ which we call \emph{spin triplet dominated} $A_{1g}$ ($B_{2g}$). Equally whether the system is in the singlet or triplet dominated regime, the form of Eqs. (\ref{s-wave}) and (\ref{d-wave}) stays the same. \newline
	
	Depending on the ratio $\alpha=J^\prime_{11}/J^\prime_{13}$ the leading pairing symmetry for low values of $\lambda_{\text{SOC}}$ is either $B_{2g}$ $d$-wave  or $A_{1g}$ $s$-wave  with accidental nodes and a sign change between inner and outer Fermi pocket. In Fig.\ref{Fig:Phasediagram}(a) we show a SOC mediated transition from $B_{2g}$ $d$-wave  to $A_{1g}$ s-wave.  For small values of $\lambda_{\text{SOC}}$ we find a nodeless $d$-wave  state where the nodes are lifted  and the gap is opened by a combination of spin singlet and triplet components with the singlet gap being dominant $|\Delta^B_1|>|\Delta^B_t|$. With increasing $\lambda_{\text{SOC}}$ the paring symmetry changes from nodeless to nodal d-wave and finally to nodeless s-wave with a dominant inter-band triplet component $|\Delta^A_t|>|\Delta^A_{1}|,|\Delta^A_{3}|$. In Fig.\ref{Fig:Phasediagram}(b) we choose parameters such that the initial state at small $\lambda_{\text{SOC}}$ is $A_{1g}$ $s$-wave and has a dominant singlet gap $|\Delta^A_1|>|\Delta^A_t|$. Increasing $\lambda_{\text{SOC}}$ lifts the nodes at the Fermi level and drives the system in an $A_{1g}$ $s$-wave  symmetry with a dominant triplet gap $|\Delta^A_t|>|\Delta^A_{1}|,|\Delta^A_{3}|$. 
	We find that for large inter-band SOC the triplet dominated $s$-wave state wins over triplet dominated $d$-wave. 
	The presence of an intra-band SOC term $\lambda_z$ does not change the phase diagram qualitatively. However, in the region where $\lambda_{\text{SOC}}$ is strong superconductivity in the $d$-wave  channel is suppressed if $\lambda_z>0$ which further stabilizes the $A_{1g}$-wave  solution. Moreover, a large triplet gap, in addition to SOC, lifts accidental nodes leading to a nodeless $s^{\pm}$ pairing symmetry.\newline
	
	Figure\ref{Fig:Phasediagram}(a) is consistent with the phase diagram of phenomenological model in Ref. \onlinecite{Khodas12} with several important differences. In particular, in our  model the singlet dominated d-wave state competes with singlet dominated bonding anti-bonding $s^\pm$ which arises from pair hopping between $xz(yz)$ and $xy$ orbitals.  Furthermore, with increasing spin-orbit coupling strength the $A_{1g}$ state contains dominant interband spin triplet component which  was absent in the simplified analysis of Ref. \onlinecite{Khodas12} as its pairing channel turned out to be strongly repulsive in the two band model. Since SOC couples even parity inter-band triplet to intra-band even parity spin singlet pairing, an attraction in the former induces a gap in the latter channel and vice versa. Consequently a gap at the FS opens even though the $E_g$ triplet is inter-band. A similar paring state with attraction in the triplet channel was recently proposed for highly doped systems with only hole pockets\cite{Vafek2017}.

	\section{Quasiparticle interference}
	
		\subsection{Local density of states in presence of impurities}

	In order to investigate the QPI signatures of the possible pairing states we need to calculate the local density of states (LDOS) for a multi-orbital system in the superconducting state. We further investigate corrections to the LDOS which arise from scattering at single charge impurities to make a statement whether there is a sign-change between inner and outer electron pocket using HAEM's method. Moreover, we consider scattering from a single magnetic impurity to reveal information about the spin structure in the system.
	
	In order to describe superconductivity and spin-resolved STM we introduce the 16 component Balian-Werthammer spinor
	
	\begin{align}
	\Psi^\dagger_\kk=\left(\Psi^\dagger_{M,\ua}(\kk), \Psi^\dagger_{M,\da}(\kk), \Psi^T_{M,\da}(-\kk), -\Psi^T_{M,\ua}(-\kk)\right)\label{BWHSpinor}.
	\end{align}
	Within this basis the Hamiltonian in the superconducting state is given by
	
	\begin{align}
	H=\frac{1}{2}\sum_{\mathbf{k}}\Psi^\dagger(\mathbf{k})\left(\begin{array}{cc}
	\mathcal{H(\mathbf{k})} & \hat{\Delta} \\ 
	\hat{\Delta}^\dagger & -\mathcal{H(\mathbf{k})}
	\end{array} \right)\Psi(\mathbf{k})\label{16x16BdGHamiltonian},
	\end{align}
	where the additional factor of $\frac{1}{2}$ accounts for double counting and 
	
	\begin{align}
	\mathcal{H(\mathbf{k})}=\left(\begin{array}{cccc}
	h^{\prime\ua\ua}_X(\kk) &  &  & \Lambda \\ 
	& h^{\prime\ua\ua}_Y(\kk) & -\Lambda^T &  \\ 
	& -\Lambda* & h^{\prime\da\da}_X(\kk) &  \\ 
	\Lambda^\dagger &  &  & h^{\prime\da\da}_Y(\kk)
	\end{array} \right).
	\end{align}
	
From Eq.(\ref{16x16BdGHamiltonian}) we find the superconducting Green's function $\hat{G}^0_\kk(\omega)=[(\omega+i\delta)\mathbbm{1}_{16\times 16}-H_{BdG}(\kk)]^{-1}$ and the
local density of states (LDOS) given by	
	
\begin{align}
\rho(\omega)&=-\frac{1}{\pi}\enspace\text{Im}\sum_{\kk}\text{Tr}\frac{\tau_1+\tau_3}{2}\hat{G}^{0}_{\kk}(\omega),
\end{align}
where we assumed sharply localized Wannier functions of the $xz$, $yz$ and $xy$ orbitals. 
	
We now introduce a single on-site non-magnetic potential scatterer to the system which is located on either sub-lattice A or B. In the one-iron unit cell it can be described as
\begin{align}
H^{1Fe}_\text{imp}(\kk,\kk^\prime)=V_{\alpha,\beta}\enspace
d^{A\dagger}_{\alpha,\sigma}(\kk)d^{A}_{\beta,\sigma}(\kk^\prime).\label{onsiteScattering}
\end{align}	We assume that the major contribution to scattering is of intra-orbital nature and project eq.(\ref{onsiteScattering}) onto the states in the two iron unit cell. Due to the fact that $d^{A/B}_{xy}$ in the projected model is a linear combination of $c_{xy^X}$ and $c_{xy^Y}$ the intra-orbital term
induces scattering between $xy^X$ and $xy^Y$ thus also contributes to inter-band scattering. In the two iron unit cell the impurity potential is given by
	
\begin{align}	H^{2Fe}_\text{imp}(\kk,\kk\prime)=\Psi_{M,\sigma}^\dagger\tilde{V}\Psi_{M,\sigma}
\label{ScatteringTerm3}
\end{align}
with 

\begin{align}
\tilde{V}=\left(\begin{array}{cccc}
	V_{yz,yz} &  &  &  \\ 
	& V_{xy^X,xy^X} &  & V_{xy^X,xy^Y} \\ 
	&  & V_{xz,xz} &  \\ 
	& V_{xy^X,xy^Y} &  & V_{xy^Y,xy^Y}
	\end{array} \right).\label{ScatteringPotential_1}
\end{align}

	Following the HAEM\cite{Hirschfeld(2015)} approach, we compute the antisymmetrized correction to the LDOS due to impurity scattering 
	
	\begin{align}
	\delta\rho^-(\omega)&= \delta\rho(\omega) - \delta\rho(-\omega), \label{drhoFouriersc}\\
	\delta\rho(\omega)&=-\frac{1}{2\pi}\text{Im}\text{Tr}\frac{\tau_0+\tau_3}{2}\sigma^0\sum_{\qv} \delta \hat{G}_\qv(\omega)
	\label{LDOS},
	\end{align}
	 with $\delta \hat{G}_{\qv}(\omega)=\sum_{\kk}\hat{G}^0_\kk(\omega)\hat{U}\hat{G}^0_{\kk+\qv}(\omega)$ being the convolution of the bare Green's functions dressed by a Nambu scattering matrix $\hat{U}=\tau_3\otimes\sigma_0\otimes\tilde{V}$ in Born approximation, i.e. with $V_{\alpha,\beta}\ll E_F$, where $E_F$ is the Fermi energy.

	The HAEM method states that the momentum integrated and antisymmetrized LDOS, $\delta\rho^-(\omega)$, qualitatively differs in case of a sign-changing OP from that of a sign-preserving one yielding a strong enhancement of the integrated response for the sign-changing but not the sign-preserving one. This method has been recently successfully applied to confirm the sign-changing nature of the order parameter in FeSe\cite{Sprau2017}, where the superconducting gaps are also extremely anisotropic.
	More recently, the method has also been applied to (Li$_{1-x}$Fe$_x$)OHFe$_{1-y}$Zn$_y$Se\cite{Du2017} where $\delta\rho ^-(\omega)$ shows a strong signal and no sign-change between 8 meV and 14 meV suggesting a sign-changing $s^\pm$ pairing symmetry. Since our previous analysis shows that most of the conclusions regarding the phase structure of the superconducting gap obtained within Born limit are robust and remain valid also well beyond this limit, we restrict our analysis to weak potential scatterers.\newline
	
	\subsection{Local density of states and phase sensitive correction to QPI}
   We would like to discuss singlet and triplet dominated $A_{1g}$- and $B_{2g}$-wave with respect to their consistency with QPI experiments in  (Li$_{1-x}$Fe$_x$)OHFe$_{1-y}$Zn$_y$Se. Recently in Ref.(\onlinecite{Myles18}) possible pairing states for intercalated FeSe have been discussed. Following Ref.[\onlinecite{Myles18}] the pairing has to obey three criteria in order to be consistent with experiments: i) fully gaped LDOS, ii)  the quasiparticle energy extrema are at or close to $k_F$ of the normal state ("back bending") and iii) two peak features in the DOS with a peak at about 8 meV and 14 meV, respectively. In the context of monolayer and intercalated FeSe it has been suggested that the second peak in the DOS arises due to pure interband gap\cite{Myles18} as in the former the gap size seen by ARPES\cite{Zhang16} (13.7 meV) significantly deviates from the peak energy (20.1 meV) seen by STM\cite{Wang12}.  This, however, does not have to be the case with the electron-intercalated materials where a gap of $13\pm2$ meV around the Fermi energy was reported\cite{Zhao16}.  Three pairing states were proposed to be consistent with experimental data. These are $A_{1g}$ $s$-wave and $B_{2g}$ $d$-wave and $E_u\otimes U(1)$ helical $p$-wave.  The $A_{1g}$ and $B_{2g}$ state were assumed to have a dominant intra-band singlet gap in order to ensure a full gap and back bending. An inter-band SOC then mixes in inter-band triplet $E_g$ pairing. For the odd parity intra-band triplet $E_u\otimes U(1)$ $p$-wave state its the odd parity inter-band singlet $A_{2u}$ that is coupled via SOC.
	
	In the following we investigate spin singlet and spin triplet dominated $A_{1g}$- and $B_{2g}$-wave  for their consistency with the citeria i)-iii) and the QPI data. For that we use Eq.(\ref{16x16BdGHamiltonian}) and calculate the LDOS ($\rho(\omega)$) and the superconducting band dispersion to ensure i)-iii) are fulfilled. In addition, for each pairing state, we calculate the antisymmetrized correction to the LDOS ($\delta\rho^-(\omega)$) to check whether the order parameter changes sign between electron pockets. Since the gaps in the intercalated FeSe are found to be isotropic\cite{Zhao16} we further present the angular dependence of the gap projected on the inner and outer electron pocket, respectively. In Appendix (\ref{sec_A2}) we also briefly discuss the QPI data for the odd parity $E_u$-wave state and show that at least within Born scattering the results are not compatible with experiment.\newline
	We find that $A_{1g}$ and $B_{2g}$ states with a dominant spin triplet gap can show back bending \emph{and}, in contrast to their singlet dominated states and the odd parity $E_u$-state, give an appropriate description of the QPI data found in experiment.
	
	One should also further note that for the HAEM method the crucial role is played by the gaps present on the Fermi surface pockets. In the present case there is also additional interband gap. We show in the Appendix C that its phase structure with respect to the gaps present on the Fermi level cannot be elucidated within phase-sensitive QPI analysis. Furthermore, we show that the sign-changing and sign-preserving gaps on the Fermi surface still determine the characteristic features of $\delta \rho^{-} (\omega)$.  \newline

	\subsection{A$_{1g}$-symmetry state}
	We start by examining spin singlet and triplet dominated $s$-wave pairing state.
	We use the singlet gaps $\Delta_1$ and $\Delta_3$ and the triplet gap $\Delta_t$ to fit band-structure and LDOS.

	\subsubsection{Spin-singlet dominated $A_{1g}$-wave state}
	
	\begin{figure}[h]
		\centering
		\includegraphics[width=8.2cm,keepaspectratio]{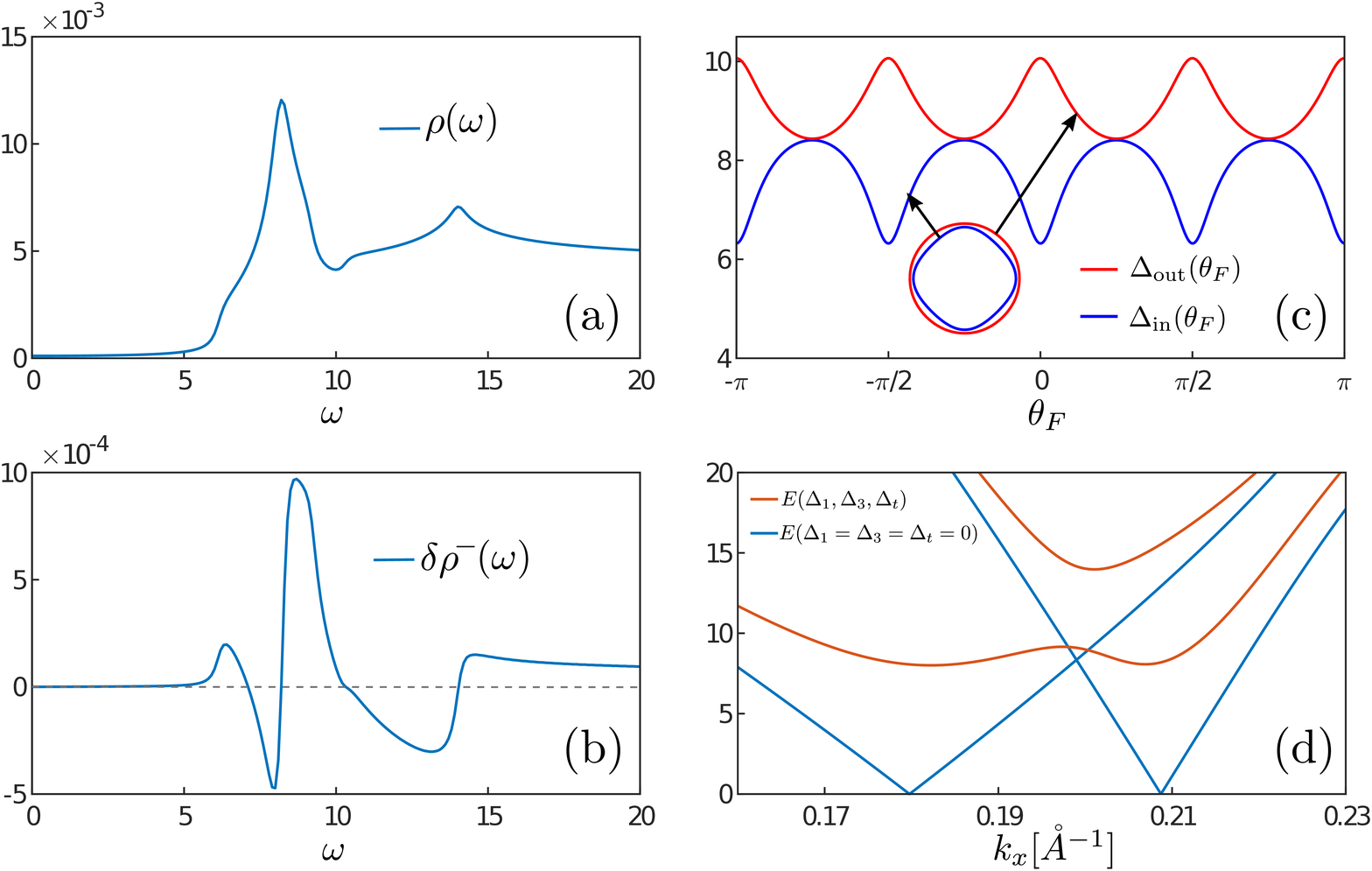}		
		\caption{Figures calculated from equation (\ref{16x16BdGHamiltonian}) for singlet dominated $A_{1g}$ and (in meV) $\epsilon_1=-45$, $\epsilon_3=-95$, $\Delta_1=10.8$, $\Delta_3=7.2$, $\Delta_t=-3$ and $\lambda_{\text{SOC}}=5$.\newline  
			(a) $\rho(\omega)$ fully gaped, with peaks at $8$ meV and $14$ meV. (b) $\delta\rho^-(\omega)$ for charge impurity in Born limit. Sign-change in $\delta\rho^-(\omega)$ between $6$ meV and $10$ meV indicates that there is no sign change of OP between inner and outer pocket. (c) Gap projected on inner (blue) and outer (red) electron pocket as function of the Fermi angle $\theta_F$. (d) Orange: Positive branches of upper and lower superconducting bands in $\Gamma M$-direction. Blue: Same bands for zero gap. The lower orange band has its minima centered above the former Fermi-level "back-bending".}	
		\label{Fig:Singlet s-wave} 
	\end{figure}

	To start with singlet dominated $s$-wave we utilize the fitting parameters presented in Ref. \onlinecite{Myles18}. As shown in Fig.\ref{Fig:Singlet s-wave}(a), $\rho(\omega)$  is indeed fully gaped and has a peak at $8$ meV and $14$ meV.  The first peak between $6-10$ meV mostly comes from the intra-band gaps at the electron pockets as can be seen by comparison with the gap projections on the inner, $\Delta_{in}$ and outer, $\Delta_{out}$, electron pocket,  Fig.\ref{Fig:Singlet s-wave}(c). At $\theta_F=\pi/4$ both gaps are equal in magnitude leading to a peak at $\omega\sim8$ meV in $\rho(\omega)$. The width of the peak is limited by 6 meV and 10 meV which  correspond to the minimum  of $\Delta_{in}$ and the maximum of $\Delta_{out}$, respectively, so that the occupied states below and above 8 meV are due to anisotropy of the gaps. The second peak at 14 meV is indeed due to inter-band pairing. It vanishes for $\lambda_{\text{SOC}}\rightarrow0$ and $\Delta_t\rightarrow0$. We find that both gaps are highly anisotropic and most important positive on both Fermi surface pockets, yielding an $s^{++}$-pairing symmetry. In Fig.\ref{Fig:Singlet s-wave}(b) $\delta\rho^-(\omega)$ exhibits a sign change in the region $6\ \text{meV}<\omega<10\ \text{meV}$ which can be interpreted as a sign preserving order parameter and thus is not compatible with the QPI data\cite{Du2017}.  Note that this state does not appear in our phase diagram, shown in Fig.2, where the A$_{1g}$ state is sign-changing.  The feature at $\omega\sim 14$ meV in $\delta\rho^-( \omega)$  for this state appears due to the small inter-band contribution and does not carry  phase information as we explain for a simple model in Appendix C. In Fig.\ref{Fig:Singlet s-wave}(d) the superconducting band dispersion in $\Gamma M$-direction ($\theta_F=\pi/4$) is shown where the minima of the lower superconducting band are located above the former Fermi level (back-bending). 
	
	It turns out that the fitting parameters for spin singlet dominated $s$-wave, proposed in Ref. \onlinecite{Myles18} would give rise to $s^{++}$ superconductivity. A possible $s^{\pm}$ pairing symmetry with a dominant spin singlet gap, on the contrary, would require $\text{sign}\Delta^A_1\neq\text{sign}\Delta^A_3$.  This  state in our anaylsis exhibits highly anisotropic gaps and possible accidental nodes (see for example  Fig.{\ref{Fig:Phasediagram}}c(1)). This makes it difficult to fit the U-shaped two peaked LDOS  in the LiOH-intercalated FeSe data without invoking a spin triplet component, induced by spin-orbit coupling that lifts the nodes.

	\subsubsection{Spin-triplet dominated $A_{1g}$-wave state}
\begin{figure}[h]
		\centering
		\includegraphics[width=8.2cm,keepaspectratio]{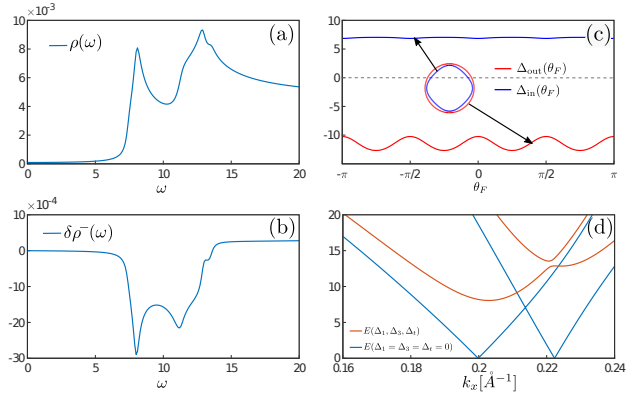}		
		\caption{Figures calculated from equation (\ref{16x16BdGHamiltonian}) triplet dominated $A_{1g}$ and (in meV) $\epsilon_1=-55$, $\epsilon_3=-105$, $\Delta_1=6.75$, $\Delta_3=-6.03$, $\Delta_t=10.27$ and $\lambda_{\text{SOC}}=7.5$.    
			(a) refer to $\rho(\omega)$ fully gaped , with peaks at $8$ meV and $13.7$ meV. (b) $\delta\rho^-(\omega)$ for charge impurity in Born limit. No sign-change of $\delta\rho^-(\omega)$ between $8$ meV and $13.7$ meV indicates an OP changing sign between inner and outer pocket. (c) Gap projected on inner (blue) and outer (red) electron pocket as function of the Fermi angle $\theta_F$. (d) Orange: Positive branches of upper and lower superconducting bands, in $\Gamma M$-direction. Blue: Same bands for zero gap . The lower orange band has its minima centered above the former Fermi-level "back-bending". }	
		\label{Fig:Triplet s-wave} 
	\end{figure}

	In Fig.\ref{Fig:Triplet s-wave}(a) and Fig.\ref{Fig:Triplet s-wave}(d) the LDOS and the superconducting band dispersion are shown. The LDOS shows a full gap and a peak at 8 meV and 13.7 meV, respectively, while the lower superconducting band shows back bending. The band projected gaps $\Delta_{in}$ and $\Delta_{out}$ we plot in Fig.\ref{Fig:Triplet s-wave}(c). Possible accidental nodes are lifted by a large $\Delta_t$
	leading to a nodeless $s^\pm$ pairing symmetry with two almost isotropic gaps. In Fig.\ref{Fig:Triplet s-wave}(b) we present $\delta\rho^-(\omega)$ which in contrast to the singlet dominated case does not change sign between 8 meV and 14 meV. If we compare Fig.\ref{Fig:Triplet s-wave}(a) and Fig.\ref{Fig:Triplet s-wave}(c) we find that the peak positions of the two peaks roughly agree with the the magnitudes of $\Delta_{in}$ and $\Delta_{out}$ and thus $\delta\rho^-(\omega)$ and the behavior of $\delta \rho^{-} (\omega)$ agrees with the experimental one. The first peak is sharper than the second since $\Delta_{in}$ is very isotropic. The second peak is broader as a consequence of an anisotropic $\Delta_{out}$ and the contribution from inter-band pairing. In contrast to the singlet dominated case $\delta\rho ^-(\omega)$ exhibits well pronounced negative peaks at $\omega_1\approx\Delta_{out}$ and $\omega_2\approx\Delta_{in}$ without a sign change between them indicating an OP that changes sign between the electron pockets consistent with the experimental QPI data and ARPES data \cite{Zhao16,Du2017}.  Also  note  that for  the spin triplet dominated $A_{1g}$ state the direct gap feature caused by $\Delta_{out}$ is very close to the interband feature in the DOS which agrees with the findings in Ref.\onlinecite{Zhao16} and \onlinecite{Du2017}.

	\subsection{$B_{2g}$-symmetry state}
	We now discuss the singlet and triplet dominated $B_{2g}$ $d$-wave pairing state.
	We use the singlet gaps $\Delta_1$ and $\Delta_3$ and the triplet gap $\Delta_t$ to fit band structure and LDOS. Note that even though our simple mean field approach leads to $\Delta_3=0$ for the $d$-wave case,  as it involves decoupling of the repulsive intraorbital Hubbard interaction, a nonzero $\Delta_3$ is necessary to achieve qualitative agreement when fitting the experimental data. In theoretical calculations, a non-zero $\Delta_3$ appears due to inclusion of the   spin fluctuation diagrams in the Cooper-pairing channel yielding momentum dependent interaction, see Ref. \onlinecite{Meier11}.

	\subsubsection{Spin-singlet dominated $B_{2g}$-state}
	
\begin{figure}[h]
		\centering
		\includegraphics[width=8.2cm,keepaspectratio]{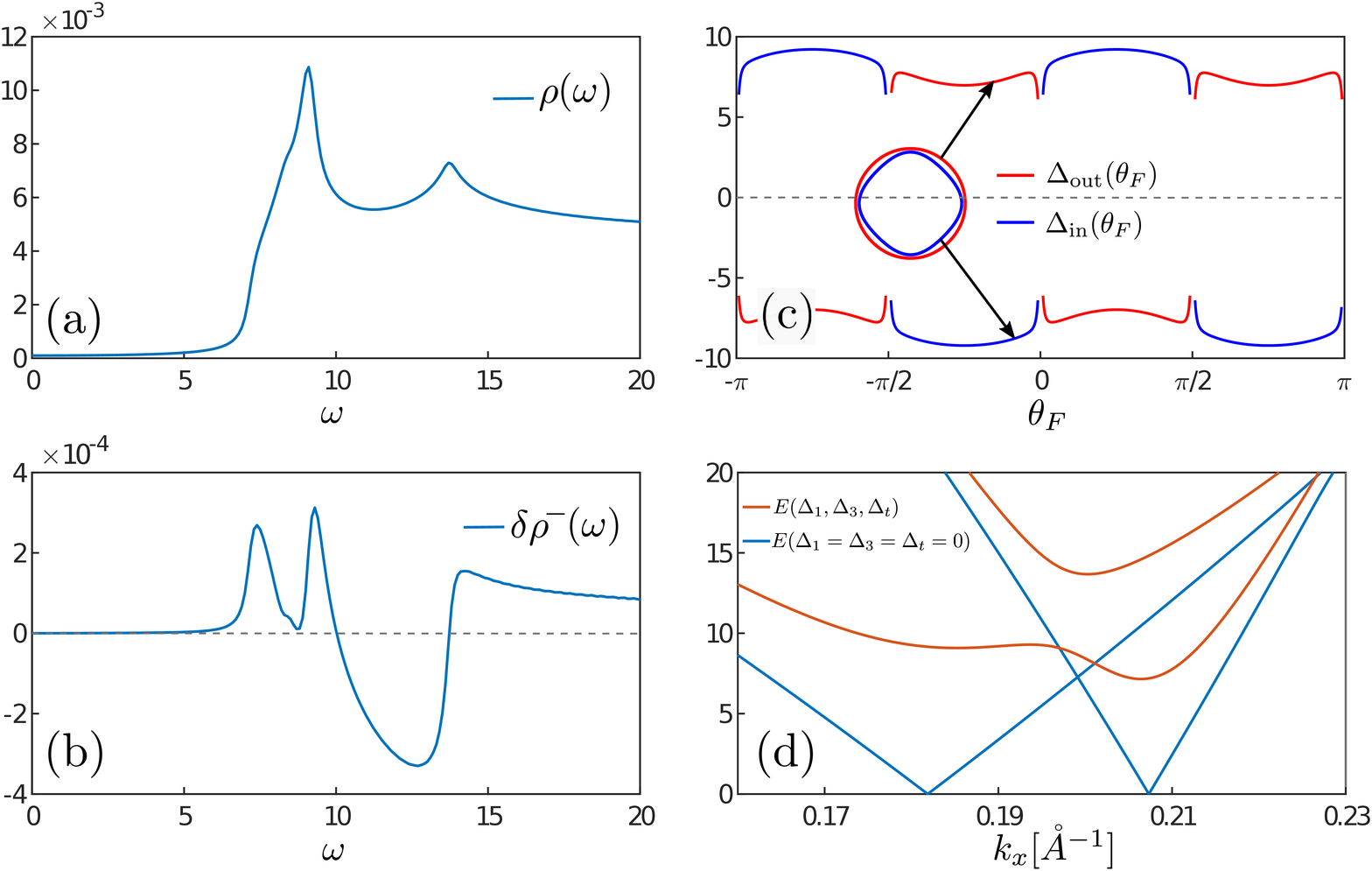}		
		\caption{Figures calculated from equation (\ref{16x16BdGHamiltonian}) for singlet dominated $B_{2g}$ and (in meV) $\epsilon_1=-45$, $\epsilon_3=95$, $\Delta_1=10.8$, $\Delta_3=7.2$, $\Delta_t=-4.8$ and $\lambda_{\text{SOC}}=0.5$.\newline     
			(a) $\rho(\omega)$ fully gaped with two peaks at $8$ meV and $14$ meV. (b) $\delta\rho^-(\omega)$ for charge impurity in Born limit. No sign-change between $6$ meV and $10$ meV indicates OP changes sign between inner and outer pocket. (c) Gap projected on inner (blue) and outer (red) electron pocket as function of the Fermi angle $\theta_F$. (d) Orange: Positive branches of upper and lower superconducting bands, in $\Gamma M$-direction. Blue: Same bands for zero gap . The lower orange band has its minima centered above the former Fermi-level "back-bending".}	
		\label{Fig:singlet d-wave} 
	\end{figure}

	For the spin singlet dominated $B_{2g}$-wave state appropriate fitting parameters were found in Ref.(\onlinecite{Myles18}). As shown in Fig.\ref{Fig:singlet d-wave}(a) and Fig.\ref{Fig:singlet d-wave}(d) the LDOS  is fully gaped and has a peak at $8$ meV and $14$ meV, moreover, the lower superconducting band shows back bending. $\Delta_{in}$ and $\Delta_{out}$ are plotted in Fig.\ref{Fig:singlet d-wave}(c). Both gaps are almost equal in magnitude and nodeless due to smallness of hybridization between the electron pockets, i.e $|\Delta_{in/out}|>\lambda_{\text{SOC}}$. The first peak in the LDOS is due to the intra-band gaps $\Delta_{in}$ and $\Delta_{out}$.  We calculate $\delta\rho^-(\omega)$ and present the results in Fig.\ref{Fig:singlet d-wave}(b).  $\delta\rho^-(\omega)$ has  positive peaks at $\sim 7$ mev and $\sim9$ meV with no sign change between them indicating the sign change between $\Delta_{in}$ and $\Delta_{out}$. Yet the third peak at $\sim 14 $ meV comes from inter-band pairing and causes two sign changes between second and third peak which is at odds with the experimental data\cite{Du2017}.

	\subsubsection{Spin-triplet dominated $B_{2g}$-state}
	Even though a triplet dominated $d$-wave scenario is not realized in the phase diagram we quickly discuss this case with respect to criteria i)-iii) and QPI data.\newline
	
\begin{figure}[h]
		\centering
		\includegraphics[width=8.2cm,keepaspectratio]{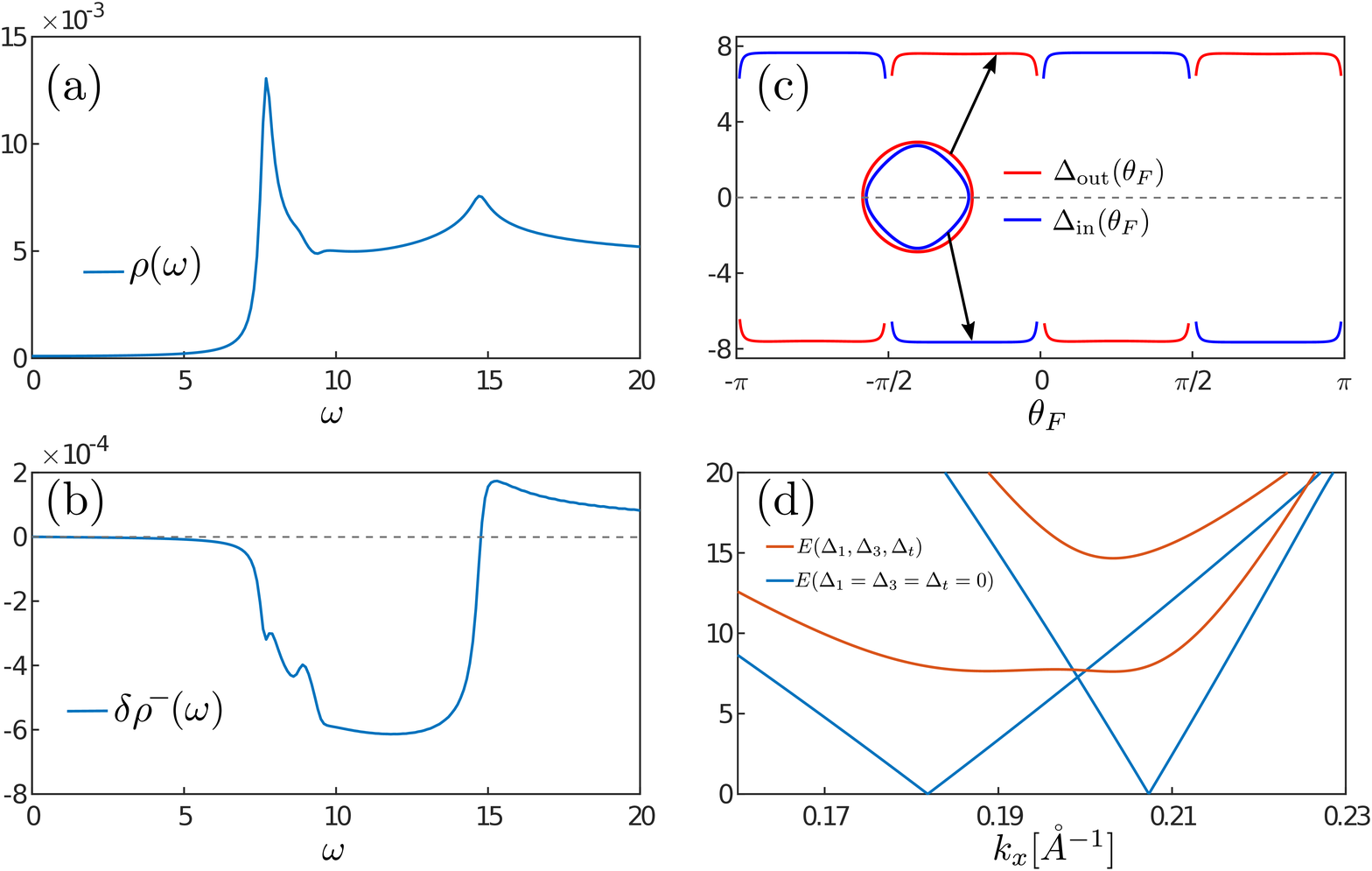}		
		\caption{Figures calculated from equation (\ref{16x16BdGHamiltonian}) for triplet dominated $B_{2g}$ and (in meV) $\epsilon_1=-45$, $\epsilon_3=95$, $\Delta_1=7.65$, $\Delta_3=7.2$, $\Delta_t=9.37$ and $\lambda_{\text{SOC}}=0.5$.\newline 
			(a) $\rho(\omega)$ fully gaped with two peaks at $8$ meV and $14$ meV. (b) $\delta\rho^-(\omega)$ for charge impurity in Born limit. No sign-change between $6$ meV and $8$ meV indicates OP changes sign between inner and outer pocket. (c) Gap projected on inner (blue) and outer (red) electron pocket as function of the Fermi angle $\theta_F$. (d) Orange: Positive branches of upper and lower superconducting bands, in $\Gamma M$-direction. Blue: Same bands for zero gap . The lower orange band has its minima centered above the former Fermi-level "back-bending". }			\label{Fig:triplet d-wave}
	\end{figure}

	As shown in Fig.\ref{Fig:triplet d-wave}(a) and Fig.\ref{Fig:triplet d-wave}(d) the LDOS  is fully gaped and has a peak at $8$ meV and $14$ meV, moreover, the lower superconducting band shows back bending. $\Delta_{in}$ and $\Delta_{out}$ are plotted in Fig.\ref{Fig:triplet d-wave}(c). Both gaps are  equal in magnitude and nodeless due to smallness of hybridization between the electron pockets. In contrast to the singlet dominated case for the triplet dominated case we use $|\Delta^B_t|>|\Delta^B_{1/3}|$ and $\text{sign}(\Delta^B_t)=\text{sign}(\Delta^B_1)=\text{sign}(\Delta^B_2)$. In Fig.\ref{Fig:triplet d-wave}(b) we plot $\delta\rho^-(\omega)$. Two small peaks between 7 meV and 8 meV reflect $\Delta_{in}$ and $\Delta_{out}$ contributions to $\delta\rho^-(\omega)$. The peak at 15 meV is due to the large inter-band gap. Overall, the behavior of $\delta\rho^-(\omega)$ is nearly consistent with experimental data as $\delta\rho^-(\omega)$ does not change sign in the experimentally relevant energy range. However, one has to bear in mind that in contrast to the triplet driven $A_{1g}$ case  the phase structure of the gaps on the Fermi surface affect the behavior of $\delta\rho^-(\omega)$ only near the lower peak, and the structure near the second (interband) peak is determined by the interband gap and does not necessarily bear information about the signs of the order parameters (see Appendix C.2).
	\newline
	As we clearly see, to obtain the agreement with QPI experiments\cite{Du2017} the spin-triplet interband component from $E_g$ state is necessarily required in both symmetry states and triplet dominated $A_{1g}$ state appears most likely one. This poses an important question how to detect it.  To stay within QPI we propose to employ spin-resolved STM as an additional tool to further specify the underlying pairing state and also to distinguish between $B_{2g}$ and $A_{1g}$-states.

	\subsection{Spin resolved STM}
In the previous section we  investigated corrections to the LDOS which arise from scattering on a single non-magnetic charge impurity and found the triplet dominated $A_{1g}$ and $B_{2g}$ pairing states in the presence of SOC to be consistent with the available experiments.  As discussed in the introduction, however, the more conventional spin singlet $A_{1g}$ and $B_{2g}$ states have also been proposed. This raises the question whether SOC effects and the spin triplet order parameter can be further verified in experiment. Here we propose spin-resolved QPI as an additional tool to further specify the underlying pairing state. To illustrate the possible capabilities of this technique, we study the corrections to the LDOS in Born approximation from a single magnetic impurity which, within our basis of eq.(\ref{BWHSpinor}), is given by $\hat{U}=\tau_3\otimes\sigma^z\otimes\hat{V}$. Here $\tau$ and $\sigma$ matrices act on Nambu and spin-space, respectively, and the matrix $\hat{V}$ describes orbital scattering. The Fourier transform of the spin resolved $\sigma^i$-projected correction to the LDOS is given by [\onlinecite{Hofmann2013}]

\begin{align}
\delta\rho_{\sigma^i}(\qv,\omega)&=-\frac{1}{2\pi i}\text{Tr}\frac{\tau_0+\tau_3}{2}\sigma^i\left[\delta\hat{G}_{\qv}(\omega)-\delta\hat{G}^*_{-\qv}(\omega)\right]\nn\\
&=-\frac{1}{2\pi}\text{Tr}\frac{\tau_0+\tau_3}{2}\sigma^i\sum_{\kk}\Bigg[\text{Im}\left[G_{\kk}(\omega)\hat{U}G_{\kk+\qv}(\omega)+G_{-\kk}(\omega)\hat{U}G_{-\kk-\qv}(\omega)\right]\nn\\
&-i\text{Re}\left[G_{\kk}(\omega)\hat{U}G_{\kk+\qv}(\omega)-G_{-\kk}(\omega)\hat{U}G_{-\kk-\qv}(\omega)\right]\Bigg]
\label{Eq:SRQPI},
\end{align}
with the real and the imaginary part being even and odd in $\qv$, respectively. Note that a non-vanishing imaginary part requires the scattering potential to break inversion symmetry, which, as we show below can result from interorbital scattering. The scattering matrix $\hat{V}=\hat{V}_{z}+\hat{V}_{x}$ can be written as

\begin{align}
\hat{V}=\left(\begin{array}{cccc}
V^{\text{intra}}_{yz,yz} &V^{\text{inter}}_{yz,xy^X}  &  &  \\
V^{\text{inter}}_{yz,xy^X}& V^{\text{intra}}_{xy^X,xy^X} &  &  \\
&  & V^{\text{intra}}_{xz,xz} & V^{\text{inter}}_{xz,xy^Y} \\
&  &V^{\text{inter}}_{xz,xy^Y}  & V^{\text{intra}}_{xy^Y,xy^Y}
\end{array} \right)+
\left(\begin{array}{cccc}
&  &V^{\text{inter}}_{yz,xz}  &V^{\text{inter}}_{yz,xy^Y}  \\
&  &V^{\text{inter}}_{xy^X,yz}  &V^{\text{intra}}_{xy^X,xy^Y}  \\
V^{\text{inter}}_{yz,xz}&  V^{\text{inter}}_{xy^X,xz}&  &  \\
V^{\text{inter}}_{yz,xy^Y}&V^{\text{intra}}_{xy^X,xy^Y}  &  &
\end{array} \right).\label{ScatteringPotential_2}
\end{align}
where, in Born approximation, $\hat{V}_{z}$ potentially contributes to the $\sigma^z$- and $\hat{V}_{x}$  to the $\sigma^{x(y)}$-polarized LDOS when taking the trace in eq.(\ref{Eq:SRQPI}). The labels intra and inter denote intra- and interorbital scattering, respectively. Note that in each block of the matrices $\hat{V}_{z}$ ($\hat{V}_{x}$) it is the off-diagonal (diagonal) elements that break inversion symmetry since under inversion $\Psi_X(-\kk)=\sigma^z\Psi_X(\kk)$ and  $\Psi_Y(-\kk)=-\sigma^z\Psi_Y(\kk)$. Those elements contribute to $\text{Im}\delta\rho_{\sigma^i}(\qv,\omega)$, while inversion symmetric ones contribute to $\text{Re}\delta\rho_{\sigma^i}(\qv,\omega)$.\newline

\begin{figure}[h]
	\centering
	\includegraphics[width=16.4cm,keepaspectratio]{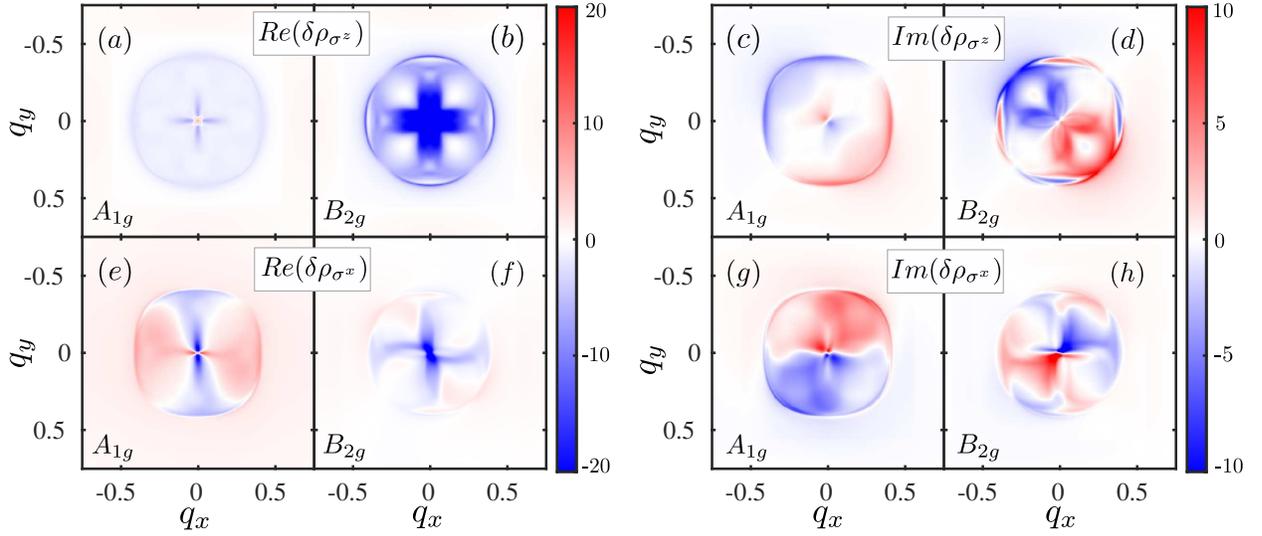}		
	\caption{Spin polarized correction to the LDOS for a magnetic impurity in Born limit.  Real and imaginary part of $\delta\rho_{\sigma^z}(\qv,\omega)$ (a-d) and  $\delta\rho_{\sigma^x}(\qv,\omega)$ (e-h) for triplet dominated $A_{1g}$ and $B_{2g}$ state at $\omega=8$ meV.}	
	\label{Fig:SRSTM}
\end{figure}

Results are presented in Fig.\ref{Fig:SRSTM} exemplary for $\omega=8$ meV where we show real and imaginary part of $\delta\rho_{\sigma^i}(\qv,\omega)$ as a function of the scattering vector $\qv$ for a $\sigma^z$ impurity with $i=z$ (a-d) and
$i=x$ (e-h). Note that a similar behavior of these quantities is found for all frequencies within the range of the two maxima in the LDOS at 8meV and 14meV, respectively.'

In Fig.\ref{Fig:SRSTM}.(a,b) $\text{Re}\delta\rho_{\sigma^z}$ shows an even in $\qv$ $C_4$-symmetric QPI pattern qualitatively similar for the $A_{1g}$ and $B_{2g}$ pairing state. In contrast to that, see Fig.\ref{Fig:SRSTM}.(c,d), $\text{Im}\delta\rho_{\sigma^z}$ is odd in $\qv$ and exhibit a few qualitative differences between $A_{1g}$ and $B_{2g}$ especially along the intensity edges. Note, however, that $\text{Im}\delta\rho_{\sigma^z}$ is caused by the inter-orbital terms in $\hat{V}_{z}$ and hence is expected to show a weakened intensity if $V^{\text{inter}}\ll V^{\text{intra}}$.

In Fig.\ref{Fig:SRSTM}.(e-h) we present real and imaginary part of  the $\sigma^x$-polarized QPI pattern $\delta\rho_{\sigma^x}(\qv,\omega)$. We would like to stress that this quantity is non-zero only in presence of either a triplet gap $\Delta^{A(B)}_t$ or SOC, which then itself induces a triplet gap. The same is true for the $\sigma^y$- polarized state which is related to $\delta\rho_{\sigma^x}(\qv,\omega)$ by $90^\circ$ rotation.

In Fig.\ref{Fig:SRSTM}(e-f) we show the real part $\delta\rho_{\sigma^x}$ which is $C_2$-symmetric and even in $\qv$ while the imaginary part is odd, see Fig.\ref{Fig:SRSTM}(g,h).
In contrast to $\delta\rho_{\sigma^z}$ the patterns for $A_{1g}$ and $B_{2g}$ representation show qualitative differences in both real and imaginary part. Moreover, if the impurity occupies one Fe site (either on sub lattice A or B) $\hat{V}_{x}$ contains scattering between $xy^X$ and $xy^Y$ components which breaks inversion symmetry and according to eq.(\ref{Eq:Projection3}) is of intra-orbital nature. Hence, we expect the QPI signal of $\text{Im}\delta\rho_{\sigma^x}$ to be of the same order of magnitude as $\text{Re}\delta\rho_{\sigma^z}$ and hence may provide a quantity that allows to distinguish triplet driven $A_{1g}$ from $B_{2g}$.

Thus we have shown that the spin-resolved QPI is a useful tool to study the details of the spin-orbit coupling and the triplet order parameter; our results suggest that a combination of spin-resolved QPI with the more conventional probes can be used to determine the symmetry as well as spin structure of pairing in iron-based superconductors.

	\section{Conclusion\label{concl}}

  In conclusion, we remind the reader that the highly electron-doped Fe-chalcogenides, in particular the  (Li$_{1-x}$Fe$_x$)OHFeSe system,  
	have been discussed intensively in the context of the standard model of spin-fluctuation induced spin singlet pairing in the limit of weak spin-orbit coupling.  In this case, the nodeless $d$-wave ($B_{2g}$) and bonding-antibonding $s_\pm$ ($A_{1g}$) spin singlet states on the electron pockets have been considered most favorable, with the additional possibility of incipient $s_\pm$ pairing via coupling to the incipient hole pocket at $\Gamma$.  
	One of these states may eventually prove to be the correct pairing state for these materials, if the effective attraction in these channels generated by spin fluctuations driven by interpocket repulsion dominate the pair vertex\cite{Meier11}.  
	
	Here, we have instead studied the new possibility raised previously\cite{Cvetkovic13}, namely that interorbital triplet components generated by intrinsic attractions possible only in the presence of spin orbit coupling are responsible for the pairing in these systems.  We explored the various
	possible Cooper-pairing symmetries using a mean-field decomposition of the Hubbard-Kanamori Hamiltonian including $A_{1g}$ and  $B_{2g}$  states.  For nonzero spin-orbit coupling, the superconducting order parameter is  a combination of  spin singlet and  spin triplet gaps in each state. Treating attraction in singlet and triplet channels on equal footing and solving the selconsitency equations, we found for weak spin-orbit coupling a dominant spin singlet and small spin triplet gap yielding a state essentially equivalent to those identified in the usual spin fluctuation approach.  For stronger spin-orbit coupling, however, the superconducting order parameter is  a combination of spin singlet and dominant spin triplet gaps in each state. Focusing on the (Li$_{1-x}$Fe$_x$)OHFeSe system, we identified the even parity $A_{1g}$-  and $B_{2g}$- pairing states with a dominant  spin triplet component to be consistent with  available experiments, including current quasiparticle interference data, whereby according to our phase-diagram the $A_{1g}$ state is slightly favored. The spin-singlet dominated $A_{1g}$ and $B_{2g}$-states in this scenario without strong spin fluctuations are not consistent with at least one of the existing experiments.
	
	In summary, to obtain a	full moderately anisotropic gap on the Fermi pockets and its  sign-changing character in agreement with experimental results on (Li$_{1-x}$Fe$_x$)OHFeSe, we 
	require {\it either} the traditional intraband $A_{1g}$ or $B_{2g}$ states generated by spin fluctuations, {\it or} the new triplet interband pair states in the same symmetry channels generated by intrinsic attraction in multiorbital correlated models.  It is clearly of interest to identify experimental tests to distinguish between these possibilities.  To this end we proposed using spin-polarized QPI to identify the possible triplet components present in the more exotic alternative states, and presented results for each of the triplet dominated states.  A clear identification of triplet interband pairing using these results would be an important step forward in understanding the unusual superconductivity in the Fe chalcogenides.\newline

\vskip .4cm
	\begin{acknowledgments}
		We thank P. M. Eugenio, O. Vafek, A. V. Chubukov  and P. Coleman for useful discussions.
		J.B. and I.E. were supported by the
		joint DFG-ANR Project (ER 463/8-1) and DAAD PPP USA
		N57316180.  P.J.H. was supported by the U.S. Department of
        Energy under Grant No. DE-FG02-05ER46236. P.A.V. acknowledges the support by the Rutgers University Center for Materials Theory Postdoctoral fellowship.
	\end{acknowledgments}

	\bibliography{Bibliography}

	\appendix
	
	\begin{widetext}

		\section{Mean-field order parameters}
		Using mean-field we decouple eq.(\ref{Interaction}) into $A_{1g}$, $B_{2g}$ and $E_g$ pairing channels. The pairing terms for $A_{1g}$- and $B_{2g}$-wave states read
		
		\begin{align}
		\Delta^A_1&=\sum_{\qv}\frac{U+J^\prime_{11}}{2}\braket{\Psi^T_{1\da}(-\qv)\tau_0\Psi_{1\ua}(\qv)}+J^\prime_{13}\braket{\Psi^T_{3\da}(-\qv)\tau_0\Psi_{3\ua}(\qv)}\nn\\
		\Delta^A_3&=\sum_{\qv}U\braket{\Psi^T_{3\da}(-\qv)\tau_0\Psi_{3\ua}(\qv)}+J^\prime_{13}\braket{\Psi^T_{1\da}(-\qv)\tau_0\Psi_{1\ua}(\qv)}\nn\\
		\Delta^A_t&=-\frac{1}{4}\sum_{\qv}\frac{U_{13}^\prime-J_{13}}{2}\left[\braket{\Psi^T_{3\ua}(-\qv)\Lambda^T\Psi_{1\ua}(\qv)}+\braket{\Psi^T_{3\da}(-\qv)\Lambda^\dagger\Psi_{1\da}(\qv)}\right]\nn\\\label{Eq:s-wave}
		\end{align}
		\begin{align}
		\Delta^B_1&=\sum_{\qv}\frac{U-J^\prime_{11}}{2}\braket{\Psi^T_{1\da}(-\qv)\tau_3\Psi_{1\ua}(\qv)}\nn\\
		\Delta^B_3&=0\nn\\
		\Delta^B_t&=-\frac{1}{4}\sum_{\qv}\frac{U_{13}^\prime-J_{13}}{2}\left[\braket{\Psi^T_{3\ua}(-\qv)\Lambda^\dagger\Psi_{1\ua}(\qv)}+\braket{\Psi^T_{3\da}(-\qv)\Lambda^T\Psi_{1\da}(\qv)}\right]\nn\\\label{Eq:d-wave}
		\end{align}
	respectively.

	\end{widetext}

	\section{$E_u$-wave state\label{sec_A2}}
	The odd parity $E_u\otimes U(1)$ $p$-wave is a novel state which was proposed for intercalated FeSe in Ref. \onlinecite{Myles18}. It is a time reversal symmetric topological superconductor which is nodeless for non zero  intra-band SOC $\lambda_z$. Consequently the order parameter is sensitive to the changes in $\lambda_z$ and the values for  $\lambda_z$, $p_{z_1}$ and $p_{z_2}$ need to be adjusted to obey criteria (i)-(iii).
	These were found in Ref. \onlinecite{Myles18} and are $\lambda_z=-31\ \text{meV}$, $p_{z_1}=4179.88\ \text{meV} \mathring{A}^3 $ and $p_{z_2}=2.5p_{z_1}$. This state does not appear in our mean-field analysis. However, for completeness we also calculate the QPI and spin-resolved QPI data for the $E_u\otimes U(1)$ helical p-wave symmetry where the pairing term within the basis of eq.(\ref{BWHSpinor}) reads
	\begin{align}
	\hat{\Delta}=\left(\begin{array}{cccc}
	& -\Delta^\dagger_{XY} & -\Delta^\dagger_{X} &  \\ 
	-\Delta^T_{XY} &  &  &  -\Delta^\dagger_{Y} \\ 
	\Delta^\dagger_{X} &  &  &  
	-\Delta^T_{XY} \\ 
	&  -\Delta^\dagger_{Y} &  -\Delta^\dagger_{XY} & 
	\end{array} \right)\label{Eq:p-waveGaps}
	\end{align}
	and
	\begin{align}
	\Delta_X&=\Delta_t\left(\begin{array}{cc}
	0 &1  \\ 
	-1& 0
	\end{array} \right),\enspace  
	\Delta_{Y}=\Delta_t\left(\begin{array}{cc}
	0&i \\ 
	-i& 0
	\end{array} \right),\quad
	\Delta_{XY}=\Delta_t\left(\begin{array}{cc}
	\Delta_1&0 \\ 
	0& i\Delta_3
	\end{array} \right).
	\end{align}
	\begin{figure}[h]
		\centering
		\includegraphics[width=8.2cm,keepaspectratio]{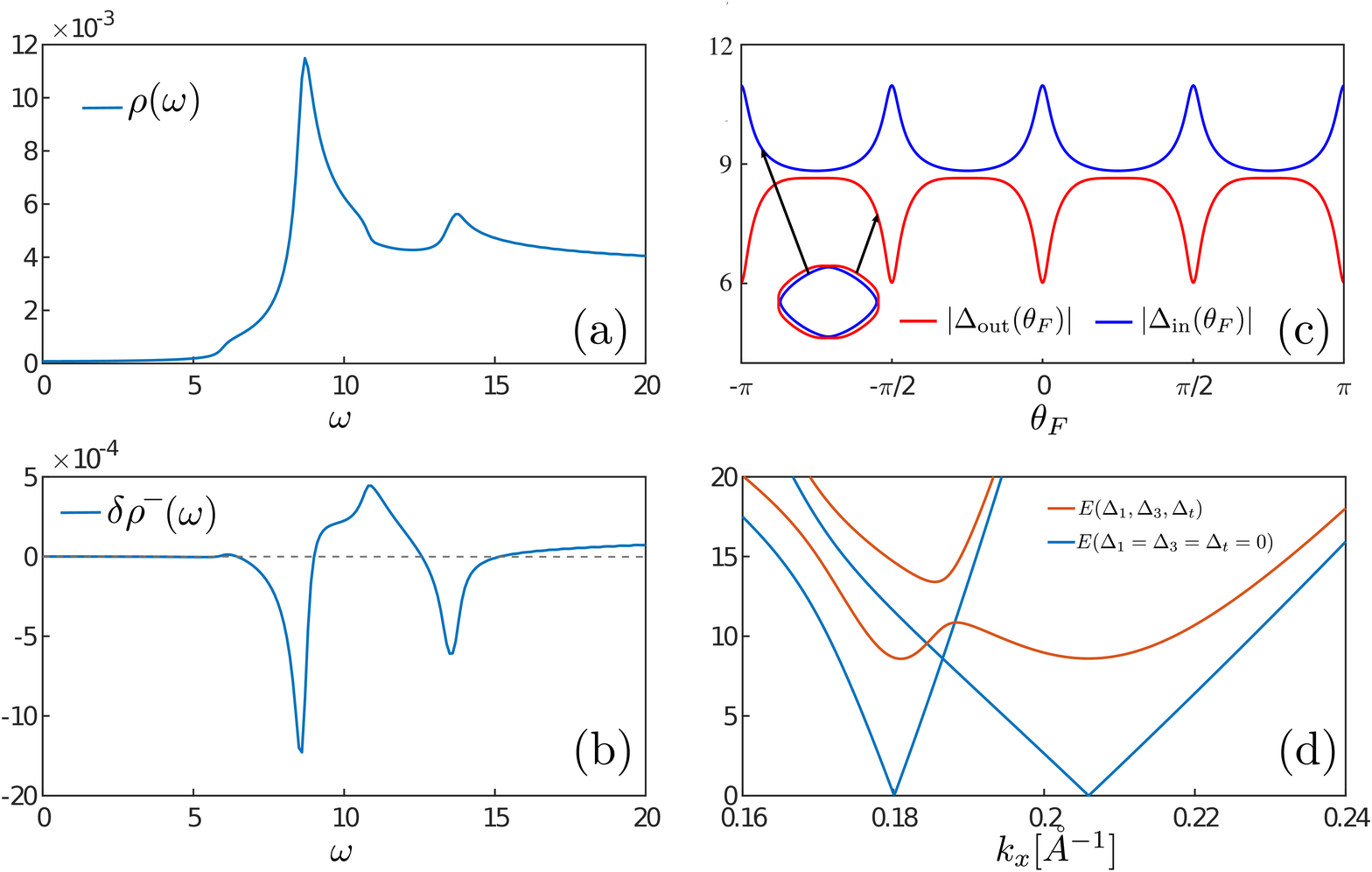}		
		\caption{Figures calculated from Eqs.(\ref{16x16BdGHamiltonian}) and (\ref{Eq:p-waveGaps}) for $E_u\otimes U(1)$ $d$-wave pairing and (in meV) $\epsilon_1=-55$, $\epsilon_3=-105$, $\Delta_1=6$, $\Delta_3=7$, $\Delta_t=9$ and $\lambda_{\text{SOC}}=3$.       
		(a) refers to $\rho(\omega)$, which  is fully gaped with two peaks at $8$ meV and $14$ meV; (b) shows $\delta\rho^-(\omega)$ for a weak (Born) charge impurity. (c) Absolute values of the superconducting gap projected on inner (blue) and outer (red) electron pockets as function of the angle of the Fermi surface, $\theta_F$. (d) Orange curves refer to positive branches of upper and lower superconducting bands along $\Gamma M$-direction. Blue curves refer to the same bands for zero superconducting gap. The low-lying energy band has its minima centered above the Fermi-level "back-bending".}	
		\label{Fig:p-wave} 
	\end{figure}

\begin{figure}[h]
		\centering
		\includegraphics[width=8.2 cm,keepaspectratio]{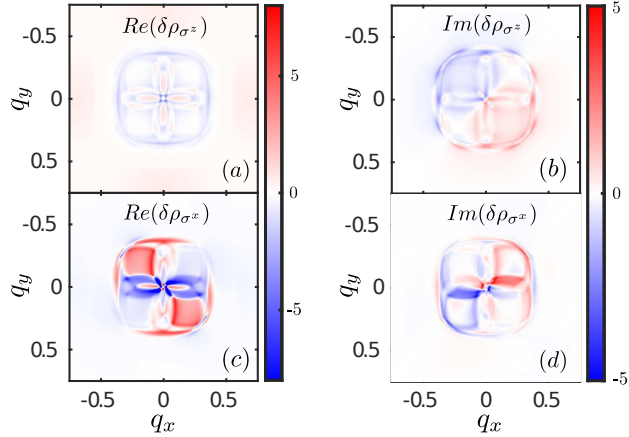}		
		\caption{Spin polarized correction to the LDOS for a magnetic impurity in Born limit.  Real and imaginary part of $\delta\rho_{\sigma^z}(\qv,\omega)$ (a,d) and  $\delta\rho_{\sigma^x}(\qv,\omega)$ (c,d) for  $E_u\otimes U(1)$p-wave state at $\omega=8$ meV.}	
		\label{Fig:SRSTMp-wave}
	\end{figure}

	In Fig.\ref{Fig:p-wave}(a) and Fig.\ref{Fig:p-wave}(d) we present the DOS and the energy band dispersion for the $p$-wave state. The  DOS show two peaks  tuned to lie  at 8 meV and 14.3 meV, respectively and the lower energy band shows the back bending. Fig. \ref{Fig:p-wave}(c) shows the absolute values of $\Delta_{in}$ and $\Delta_{out}$ projected on the Fermi surface pockets. In Fig.\ref{Fig:p-wave}(b) $\delta\rho^-(\omega)$ is shown. Here, one finds that the main feature of $E_u\otimes U(1)$ in $\delta\rho^-(\omega)$ is that there are two sign changes between the peak energies of the total DOS (i.e. two negative peaks and positive values in between them), which does not agree with experiment\cite{Du2017}. Therefore this state does not seem consistent with the QPI data at least within Born scattering limit\cite{Du2017}. For consistency the spin-resolved QPI  for the odd parity $E_u\otimes U(1)$ p-state  is presented in Fig.\ref{Fig:SRSTMp-wave}.

	\section{HAEM'S method for strong inter-band pairing\label{secA_3}}
	The SOC induced coupling between the spin singlet and triplet pairing channels translates into a coupling between intra- and inter-band order parameters in band space. It has been argued\cite{Myles18} that the second peak, seen in STM\cite{Du2017} can be at least partially due to an interband gap. In order to investigate how the predictions of HAEM's theory are affected by inter-band pairing we consider a simple model of a two band superconductor with superconductivity driven by intra- and inter-band pairing. 
	\begin{align}
	H=\sum_{\kk}
	\Psi^\dagger(\kk)
	\left(\begin{array}{cccc}
	\xi_1(\kk) &  & \Delta_{11} & \Delta_{12} \\ 
	& \xi_2(\kk) & \Delta_{21} & \Delta_{22} \\ 
	\Delta^\dagger_{11} &\Delta^\dagger_{21} & -\xi_1(\kk) &  \\ 
	\Delta^\dagger_{12} & \Delta^\dagger_{22} &  & -\xi_2(\kk)
	\end{array} \right)\Psi(\kk)\label{BdG}
	\end{align}
	where $\Psi^\dagger(\kk)=\left(c^\dagger_1(\kk)\enspace  c^\dagger_2(\kk)\enspace c_1(-\kk)\enspace  c_2(-\kk)\right)$. The band dispersions are assumed to be simple parabolic ones 
	$\xi_1(\kk)=\frac{k^2}{2m}-\mu_1$ and $\xi_2(\kk)=\frac{k^2}{2m}-\mu_2$. Hence,
	$\xi_2=\xi_1-2B$ with $B=\frac{1}{2}(\mu_2-\mu_1)>0$. We only distinguish  between intra- and inter-band pairing and neglect for a moment their spin symmetry (i.e. consider them to be spin singlet)  to get a better understanding on how the {QPI data} is affected by the inter-band pairing. We linearize both bands, and calculate $\delta\rho^-(\omega)$ and investigate how it depends on $\Delta_{12}$, the relative phase between $\Delta_{12}$ and $\Delta_{21}$ and the band offset (direct gap).
	
	\subsection{QPI for Inter-band \label{secA_3_a}}
	We start by examining $\delta\rho^-(\omega)$ for zero intra-band gaps $\Delta_{11}=\Delta_{22}=0$ and assume $|\Delta_{12}|=|\Delta_{21}|$. From the Green's function $G(\kk,\omega)^{-1}=\left[\omega+i\pmb{\delta }-H_{BdG}\right]$ we calculate the momentum integrated Green's function  as 
	\begin{widetext}
		\begin{align}
		\hat{G}(\omega)=-\pi\frac{\left(\begin{array}{cccc}
			\omega-B &  &  & \Delta_{12} \\ 
			0& 0 &0 &0  \\ 
			0&0 &0  &0  \\ 
			\Delta^\dagger_{12} & &  & \omega-B
			\end{array} \right)}{\sqrt{|\Delta|^2-(\omega-B)^2-i\pmb{\delta}\text{sign}(\omega-B)}}
		-\pi\frac{\left(\begin{array}{cccc}
			0 &  &  & 0 \\ 
			0& \omega+B &\Delta_{21} &0  \\ 
			0&\Delta^\dagger_{21} &\omega+B  &0  \\ 
			0 & &  & 0
			\end{array} \right)}{\sqrt{|\Delta|^2-(\omega+B)^2-i\pmb{\delta}\text{sign}(\omega+B)}}
		\end{align} 
	\end{widetext}
	where $\pmb{\delta}=\delta \text{sign}(\omega)$. The local LDOS we obtain from $\rho(\omega)=-\frac{\text{sign}(\omega)}{\pi}\sum_{i=1}^2\sum_{\kk}\text{Im}\hat{G}(\kk,\omega)_{ii}$.  The result is

	\begin{widetext}
		\begin{align}
		\delta\rho^-(\omega)=\text{sign}(\omega)2N_0^2\pi\text{Im}\left(\frac{2B^2-2\omega^2+\Delta_{12}\Delta_{21}^\dagger+\Delta_{12}^\dagger\Delta_{21}}{\sqrt{|\Delta|^2-(\omega-B)^2-i\pmb{\delta}\text{sign}(\omega-B)}\sqrt{|\Delta|^2-(\omega+B)^2-i\pmb{\delta}\text{sign}(\omega+B)}}\right)\label{RhoM},
		\end{align}
	\end{widetext}
	which now explicitly includes the interband pair gaps $\Delta_{12}$ and $\Delta_{21}$.
	Eq. (\ref{RhoM}) has four poles.
	\begin{align}
	\omega_{1,2}&=B\pm|\Delta_{12}|\\
	\omega_{3,4}&=\pm|\Delta_{12}|-B
	\end{align}
	For $\omega>0$ one needs to consider the cases: i) $B<|\Delta_{12}|$ and ii) $B>|\Delta_{12}|$ where in the former $\omega_1$ and $\omega_3$ and in the latter $\omega_2$ and $\omega_4$ correspond to the peak energies in $\delta\rho^-(\omega)$.\newline\newline 
	\underline{i) $B<|\Delta| $}\newline
	
\begin{figure}[h]
		\centering
		\includegraphics[width=8.2cm,keepaspectratio]{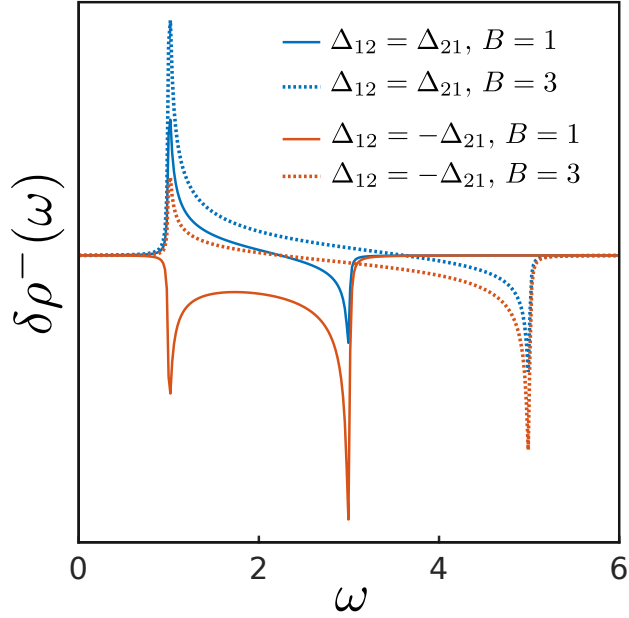}		
		\caption{$\delta\rho^-(\omega)$ in arbitrary energy units for the two cases (i) $B<|\Delta_{12}|=2$ (solid curves) and (ii) $B>|\Delta_{12}|=2$ (dotted curves). (i): QPI pattern is sensitive to phase difference showing $s^{\pm}$-pattern for $\Delta_{12}=-\Delta_{21}$ and $s^{++}$-pattern for $\Delta_{12}=\Delta_{21}$. (ii): Both $\Delta_{12}=\pm\Delta_{21}$ exhibit $s^{++}$-pattern. } 	
		\label{Fig:delta_rho_inter}
	\end{figure}

	In Fig.\ref{Fig:delta_rho_inter}, we plot $\delta\rho^-(\omega)$ in arbitrary energy units as a function of $\omega$ and  $B<|\Delta|$.
	For $\Delta_{12}=\Delta_{21}$ the behavior of $\delta\rho^{-}$ is what we define here as ``odd", meaning that between the two intraband gap 
energies $\delta\rho^-$ changes sign, so
	 the QPI-pattern is  $s^{++}$-like (solid blue curve). In case of  $\Delta_{12}=-\Delta_{21}$ $\delta\rho^-$ is ``even", and one obtains as "$s^{+-}$-like" pattern (solid orange curve). Hence, if the inter-band gap is larger than the band offset  HAEM is sensitive to the relative phase between $\Delta_{12}$ and $\Delta_{21}$.\newline\newline

	\underline{ii) $B>|\Delta|$}\newline
	
	If $B>|\Delta_{12}|$ the situation is a different one as can be seen by the dotted curves in Fig.\ref{Fig:delta_rho_inter}. In both cases $\Delta_{12}=\pm\Delta_{21}$ and $\delta\rho^-$ is ``odd", mimicking an $s^{++}$-pattern. Consequently,  HAEM's method is  \emph{not} sensitive to the relative phase  between $\Delta_{12}$ and $\Delta_{21}$.	Therefore, in a system with a large dominant inter-band gap the QPI signal depends not only on the relative phase between $\Delta_{12}$ and $\Delta_{21}$ but also on the ratio between $|\Delta|$ and $B$, hence on the band structure. 
	
		\subsection{Inter+Intra-Band \label{secA_3_b}}
\begin{figure}[h]
		\centering
		\includegraphics[width=8.2cm,keepaspectratio]{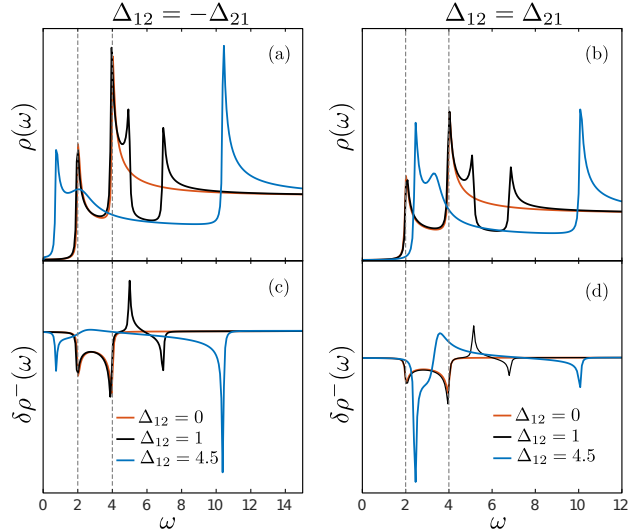}		
		\caption{$\rho(\omega)$ and $\delta\rho^-(\omega)$ as function of $\omega$ for various values of $\Delta_{12}$. Data shown for $\Delta_{12}=-\Delta_{21}$ [(a) and (c)] and $\Delta_{12}=\Delta_{21}$ [(b) and (d)], respectively. Intra-band gaps and band offset are $\Delta_{11}=2$, $\Delta_{22}=-4$ and $B=5$. Vertical dashed lines mark energies of intra-band gaps} 	
		\label{Fig:delta_rho_inter+intra}
	\end{figure}

	The influence of the inter-band gap on the QPI does affect the HAEM results on sign-changing and sign preserving intra-band gaps. To show this we numerically present the $\delta \rho^- (\omega)$  for $\Delta_{11}\neq\Delta_{22}$ and $\text{sign}(\Delta_{11})=-\text{sign}(\Delta_{22})$ leading to an initial $s^{\pm}$ pattern. Then we increase $|\Delta_{12}|$ and show that at a certain magnitude the phase information is lost.  
	The results are plotted in Fig.\ref{Fig:delta_rho_inter+intra} where in the left and right panel we have  $\Delta_{12}=-\Delta_{21}$ and $\Delta_{12}=\Delta_{21}$, respectively.		
	Our results for the LDOS are presented in  Fig.\ref{Fig:delta_rho_inter+intra}(a) and  Fig. \ref{Fig:delta_rho_inter+intra}(b).
	If $\Delta_{12}=0$, see orange curve, only intra-band gaps  $\Delta_{11}=2$ and $\Delta_{22}=-4$ are present which cause a two peak feature in the LDOS which we call peak 1 and 2. If we switch the inter-band paring $\Delta_{12}=1$ on, see black curves, peak 3 and 4 appear as a consequence of inter-band pairing, whereas peak energies 1 and 2 are only slightly affected.
	If $\Delta_{12}$ is increased further, see blue curves, peak 2 and 3 merge. 
	
	 In the absence of inter-band pairing the QPI pattern in Fig.\ref{Fig:delta_rho_inter+intra}(c)-(d) is of $s^{\pm}$ type, see orange curve. For moderate values $\Delta_{12}=1$ one can nicely distinguish between intra- and inter-band contributions Peak 1 and 2 show an $s^\pm$ pattern due to the sign change between $\Delta_1$ and $\Delta_2$, whereas  between peak 3 and 4 the pattern is an $s^{++}$  since the inter-band gap is smaller that the band offset B, see Fig.\ref{Fig:delta_rho_inter}. For moderate values of $\Delta_{12}$, see blue curves, peak energies 1 and 2 start to deviate from $\Delta_1$ and $\Delta_1$ as marked by the vertical dashed lines. As soon as peak 2 and 3 merge the $s^{\pm}$ pattern between peak 1 and peak 2 becomes more and more difficult to resolve. The blue curve corresponds to the situation where  the inter-band gap $\Delta_{12}$ is larger than the intra-band gap but smaller that the band offset.  The latter condition causes an $s^{++}$ pattern between peak 3 and 4 independent of the relative phase between  $\Delta_{12}$ and $\Delta_{21}$, sec.\ref{secA_3_a}. This is accompanied by $\Delta_{11}$ and $\Delta_{22}$ having a $s^{\pm}$ symmetry. The combination of both patterns leads to a sign-change of $\delta\rho^-(\omega)$ in the region $2<\omega<4$ which might be misinterpreted as a an $s^{++}$ pattern between $\Delta_{11}$ and  $\Delta_{22}$ and hence carries no concrete phase information.

\end{document}